\documentclass[twocolumn,prb,floatfix`]{revtex4-1}

\usepackage{graphicx}
\usepackage{epsfig}
\usepackage{verbatim}
\usepackage{bm}

\usepackage{physics}
\usepackage{times}
\usepackage{amsmath}
\usepackage{amsfonts}
\usepackage{amssymb}
\usepackage{graphicx}

\DeclareMathOperator\erfc{erfc}

\begin{document}

\title{Low overhead Clifford gates from joint measurements in surface, color, and hyperbolic codes}
\author{Ali Lavasani}
\author{Maissam Barkeshli}
\affiliation{Department of Physics, Condensed Matter Theory Center, University of Maryland, College Park, Maryland 20742, USA
and Joint Quantum Institute, University of Maryland, College Park, Maryland 20742, USA}
\begin{abstract}
One of the most promising routes towards fault-tolerant quantum computation utilizes topological quantum error correcting
codes, such as the $\mathbb{Z}_2$ surface code. Logical qubits can be encoded in a variety of ways in the surface code,
based on either boundary defects, holes, or bulk twist defects. However proposed fault-tolerant implementations of the Clifford
group in these schemes are limited and often require unnecessary overhead. For example, the Clifford phase gate in
certain planar and hole encodings has been proposed to be implemented using costly state injection and distillation protocols.
In this paper, we show that within any encoding scheme for the logical qubits, we can fault-tolerantly implement the
full Clifford group by using joint measurements involving a single appropriately encoded logical ancilla. This allows us to
provide new low overhead implementations of the full Clifford group in surface and color codes. It also provides the
first proposed implementations of the full Clifford group in hyperbolic codes. We further use our methods to propose
state-of-the art encoding schemes for small numbers of logical qubits; for example, for code distances $d = 3,5,7$, we propose
a scheme using $60, 160, 308$ (respectively) physical data qubits, which allow for the full logical Clifford group to be implemented
on two logical qubits. To our knowledge, this is the optimal proposal to date, and thus may be useful for
demonstration of fault-tolerant logical gates in small near-term quantum computers.
\end{abstract}

\maketitle


\section{Introduction}

A crucial pillar of universal fault-tolerant quantum computation is the ability to perform quantum error correction \cite{terhal2015,campbell2017}.
Schematically, given a physical qubit with an error probability $p$, a quantum error correcting code allows one to reach a target
error rate for a logical qubit with error probability $p_\text{fail} \sim (p/p_{th})^{d/2}$, where $d$ is the code distance and $p_{th}$
is the error threshold of the code\cite{Fowler2013analytic}. It is therefore desirable to implement a code which maximizes $d$ and $p_{th}$ to the extent
possible for a given set of physical resources. Codes that allow $d$ to be arbitrarily large while maintaining
local interactions between the physical qubits are \it topological error correcting codes\rm, which utilize the physics of
topological states of matter \cite{kitaev2003,wang2008,nayak2008}. In topological error correcting codes on the Euclidean plane with
local interactions, the ratio of the number of physical qubits $N_{phys}$ to the number of logical qubits $N_{L}$
scales as $N_{phys}/N_L = \mathcal{O}(d^2)$~\cite{bravyi2010tradeoffs}.

The simplest topological error correcting code is known as the $\mathbb{Z}_2$ surface code \cite{bravyi1998,dennis2002,fowler2012}, and possesses a relatively high
error threshold; for certain error models the error threshold is quoted to be $p_{th} \sim 1 \%$\cite{stephens2014fault}.
Given the rapid experimental advances in qubit technology using various physical platforms, it is reasonable to expect that the $\mathbb{Z}_2$ surface
code will play an important role in near-term demonstrations of fault-tolerance. Closely related error correcting
codes are the color codes \cite{bombin2006}  and hyperbolic codes  \cite{freedman2002,breuckmann2016,breuckmann2017}.
The color code is effectively two independent copies of the $\mathbb{Z}_2$ surface code \cite{kubica2015};
while it has a lower error threshold\cite{stephens2014efficient}, it allows for transversal implementation of Clifford gates and improves the space-time overhead \cite{landahl2014}. The
hyperbolic codes are related to the $\mathbb{Z}_2$ surface code on a tiling of hyperbolic space; they allow one to improve
the scaling of the ratio $N_{phys}/N_L$ to be independent of $d$, at the cost of requiring non-local interactions\cite{breuckmann2016}.

As we review below, logical qubits can be encoded in the surface code in a number of different ways: through (1) boundary defects,
which are domain walls between alternating boundary conditions, (2) holes, or (3) bulk twist defects. Hybrid approaches that combine
any or all of the above are also possible.

The set of fault-tolerant logical operations that can be performed using the $\mathbb{Z}_2$ surface code form the Clifford group.
In addition to Pauli operations on single qubits, this group is generated by the single qubit Hadamard gate $H$, phase gate $S$, and
two-qubit CNOT gate. A variety of methods are known for implementing these gates in the $\mathbb{Z}_2$ surface code,
however they depend sensitively on the encoding scheme \cite{dennis2002,fowler2012,horsman2012,hastings2014,litinski2017,yoder2017,brown2017}.
In particular, for the schemes that are based purely on boundary or hole defects\cite{fowler2012,horsman2012}, implementing the
Clifford phase gate require a costly state distillation protocol with a large overhead that scales exponentially with the number of
distillation rounds. The CNOT and $H$ gates in these encoding schemes also require unnecessary overhead, as we argue below.
To avoid these overhead costs, various encoding scheme specific solutions have been devised\cite{brown2017,litinski2017}, which
we will review briefly below.

In the past few years, an approach to topological quantum computation has been developed that utilizes the idea of
topological charge measurements.\cite{bonderson2009,barkeshli2016mcg} In particular, Ref. \onlinecite{barkeshli2016mcg}
demonstrated that topological charge measurements along certain `graph' operators could in principle be utilized to implement non-trivial
fault-tolerant logical unitary gates (see also Ref. \onlinecite{cong2016}).

In this paper, we demonstrate how to efficiently implement the full Clifford group with low overhead in the surface code,
using any of the above encoding schemes for the logical qubits. Our method is based on fault-tolerantly implementing the necessary
topological charge measurements using any logical encoding scheme in the surface code. Notably, this allows us to implement
the Clifford phase gate in any encoding scheme without using costly state distillation protocols, and further
allows implementation of $H$ gates and arbitrarily long-range CNOT gates with minimal overhead.

Our results can be applied not only to 2D surface codes in any logical encoding scheme, but also to 3D surface codes. We further apply our methods
to both color codes and hyperbolic codes. In the context of hyperbolic codes, we provide the first proposed implementations
of Clifford gates. In the context of color codes, we also propose novel efficient methods for implementing Clifford group operations in a hole based encoding scheme,
which provides some advantages over alternate proposals\cite{bombin2006,landahl2014,litinski2017b}.

Finally, using these insights, we further present encoding schemes for small numbers of logical qubits that allow fault-tolerant
implementation of the full Clifford group with minimal overhead in terms of number of physical qubits for a given code
distance. This leads us to state-of-the art code designs that minimize number of physical qubits while allowing
for all Clifford group operations to be implemented on two logical qubits. Specifically, for code distances $d = 3,5,7$, we propose
a scheme using $60, 160, 308$ (respectively) physical data qubits, which allow for the full logical Clifford group to be implemented
on two logical qubits. To our knowledge, this is the optimal proposal to date, and thus may be useful for near-term experiments
to demonstrate fault-tolerance.

We note that to obtain universal fault-tolerant quantum computation, the Clifford group must be supplemented with an additional gate,
such as the single qubit $\pi/8$ phase gate. In the codes that we study in this paper, this gate inevitably requires magic state injection
and distillation. In this paper we focus on efficient fault-tolerant implementations of gates in the Clifford group, and do not further consider
the $\pi/8$ phase gate.

The rest of this paper is organized as follows. In Sec. \ref{sec:rev} we provide a review of active error correction with the surface code, together with a brief review
of the different encoding schemes and proposals for carrying out quantum computation with them. In Sec. \ref{sec:circuits}, we explain the abstract
joint measurement circuits that allow implementation of the full Clifford group. In Sec. \ref{sec:surface}, we demonstrate how to implement these measurement
circuits in the surface code, using two methods: with the aid of CAT states, or using a twist defect logical ancilla. In Secs. \ref{sec:hyper}-\ref{sec:color} we further apply these
results to hyperbolic and color codes. Finally in Sec. \ref{sec:resource} we provide resource overhead estimates in terms of the number of physical qubits required
to carry out our proposal, and compare them to other existing proposals. In particular, we provide novel state-of-the-art proposals that minimize
the number of physical qubits for small numbers of logical qubits, while allowing full implementation of the Clifford group.
In Sec. \ref{sec:conclusion} we provide some concluding remarks.

\section{Review of logical qubit encodings and Clifford gates in surface code}\label{sec:rev}

We begin with a brief review of the various proposals\cite{dennis2002,fowler2012,horsman2012, hastings2014,yoder2017}
for quantum computing with the surface code.

\subsection{Planar encoding}

\begin{figure}[h]
\vspace{1cm}
\centerline{\includegraphics[width=0.5\textwidth]{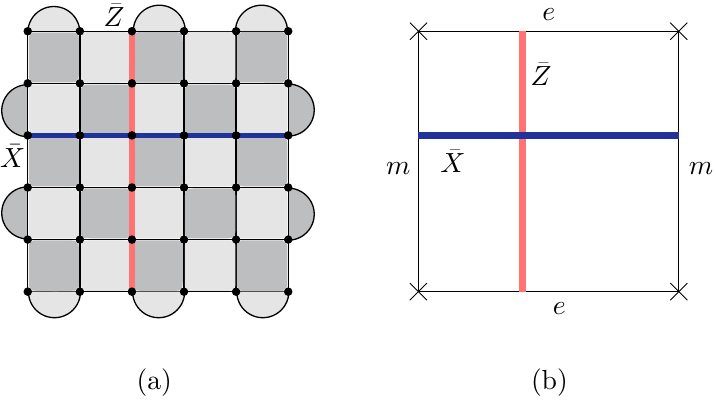}}
\caption{(Color online) a) A simple distance $6$ surface code encoding $1$ logical qubit. Black dots are physical qubits. Each dark(light) plaquette represents a $X$($Z$) stabilizer. The semicircles at the boundaries are also stabilizers that involve just two qubits. The logical $\bar X$ and $\bar Z$ operators are shown by horizontal blue (dark gray) and vertical red (light gray) strings respectively. b) Schematic diagram of the surface code in (a).}\label{fig:surfacebasic1}
\end{figure}

The simplest type of surface code is the planar code based on boundary defects.\cite{bravyi1998,freedman2001,dennis2002}
We consider a physical qubit at each site of a square lattice, as shown in  Fig. \ref{fig:surfacebasic1}a.
Each plaquette $p$ is associated with a stabilizer $S_p$, with dark plaquettes representing $X$ stabilizers and
light plaquettes representing $Z$ stabilizers:
\begin{equation}
S_p=\prod_{i\in \partial p} \sigma_i,\qquad \sigma=
  \begin{cases}
      X, & \text{if}\  p\ \text{is dark}\\
      Z, & \text{if}\ p\ \text{is light},
    \end{cases}
\end{equation}
where $\partial p$ denotes the boundary of the $p$ plaquette. In the bulk, the stabilizers have support on four physical qubits. Violations of $X$-type stabilizers are
referred to as $e$ particles, and violations of $Z$-type stabilizers are referred to as
$m$ particles. Local operators in the bulk can only create $e$ particles in pairs, and similarly for $m$ particles.

On the boundary, the stabilizers, shown as semicircles in Fig. \ref{fig:surfacebasic1}a, involve
two physical qubits. An edge with only $Z$ type stabilizers is referred to as an $e$ boundary, because applying
a $Z$ operator on an edge qubit can create a single $e$ particle; therefore, the $e$ particles are `condensed' on
such an edge. Similarly, an edge with only $X$ type stabilizers is referred to as an $m$ boundary (see Fig. \ref{fig:surfacebasic1}a). To avoid drawing the
entire lattice, we use schematic diagrams whenever possible, as shown in Fig. \ref{fig:surfacebasic1}b.
The crosses on the edges, which are domain walls between the two types of boundaries, are referred to
as boundary defects.

The stabilizers all commute with each other. The code space
$\mathcal{C}$ is defined as the set of states that are eigenvectors
of all stabilizer operators with eigenvalue $+1$:

\begin{equation}
\mathcal{C}=\qty{\ket{\psi}:\, S_p\ket{\psi}=+\ket{\psi}\quad \forall p\label{equ:stabilizer}}.
\end{equation}

The dimensionality of $\mathcal{C}$ determines how many logical qubits can be encoded in this surface code.
For the lattice shown in Fig. \ref{fig:surfacebasic1}a, there is one less stabilizer than physical qubits.
Therefore $\mathcal{C}$ is two-dimensional and corresponds to the encoded logical qubit.
It is possible to encode more than one logical qubit in one patch if one uses more defects on
the boundary; $2n$ boundary defects can be used to encode $n-1$ logical qubits. However in the
planar code, each logical qubit is associated with a separate patch.

Logical operators are associated with those unitary transformations which leave the code subspace $\mathcal{C}$ invariant, but
which act non-trivially within $\mathcal{C}$. Logical Pauli operators $\bar{Z}$ and $\bar{X}$ correspond to a tensor product
of Pauli operators for each physical qubit along a given string:
\begin{equation}
\bar Z =\prod_{i \in l} Z_i,\qquad \bar X=\prod_{i \in l'} X_i ,
\end{equation}
where $l$ and $l'$ are the light red  and dark blue strings, respectively, depicted in Fig. \ref{fig:surfacebasic1}a.
The choice of $l$ and $l'$ is unphysical; any string $l$ that connects the top and bottom edges
is sufficient for $\bar Z$, and analogously for $\bar X$. Physically, $\bar Z$ corresponds to an $e$
particle being transported between the top and bottom edge, while $\bar X$ corresponds to an $m$
particle being transported between the left and right edge.

The distance of the code, $d$, is the minimum number of Pauli operators that appear in a nontrivial
logical operator. Here, both $\bar Z$ and $\bar X$ have length $6$, hence it is a distance $d=6$ code.

\subsubsection{Error correction in surface codes}

Here we briefly review the proposal for active quantum error correction using the surface code.
In this approach, all of the stabilizers $S_p$, for every plaquette, are measured in each round of quantum
error correction. By constantly measuring the stabilizers $S_p$ for every plaquette, we can ensure that the state
of the system remains an eigenstate of each stabilizer.

\begin{figure}[h]
\vspace{1cm}
\centerline{\includegraphics[width=0.4\textwidth]{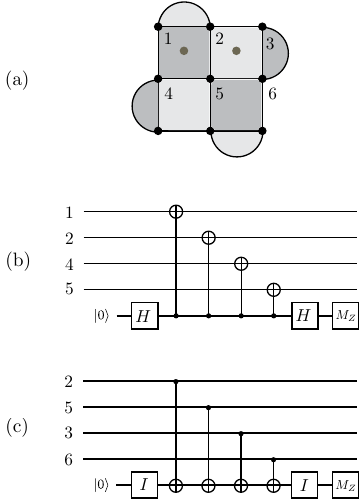}}
\caption{a) A distance $3$ planar code encoding one logical qubit. The gray dots in the center of plaquettes are the syndrome qubits for the corresponding stabilizer. b) \& c)  are quantum circuits used to measure a typical $X$ and $Z$ stabilizer respectively. Numbers represent physical qubits shown in (a).}\label{fig:surfaceStb}
\end{figure}

While each $S_p$ is a physical operator on four qubits, it can be measured using only
two-qubit CNOT operations with the aid of a physical ancilla qubit, which can be placed at the center
of each plaquette. To measure a stabilizer, such as $X_1 X_2 X_3 X_4$, one can use the circuit shown
in Fig. \ref{fig:surfaceStb}b \cite{fowler2012}. The extra ancilla qubit used in this circuit is called
the syndrome or measurement qubit. In Fig. \ref{fig:surfaceStb}c, a similar circuit is shown
which is used to measure a typical $Z$ stabilizer.



\begin{figure}[h]
 \vspace{1cm}
 \centerline{\includegraphics[width=0.3\textwidth]{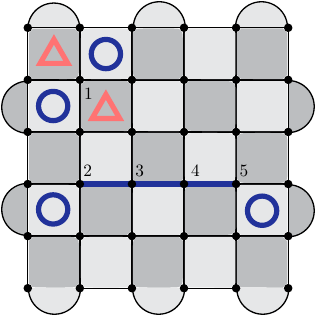}}
 \caption{(Color online) Error syndromes can be used to detect and correct errors. A single bit flip error on qubit $1$ will create two neighboring $m$ particles indicated here by blue (dark gray) circles whereas a phase flip error will create two $e$ particles (shown as red (light gray) triangles). Error strings like the one shown at the bottom, can create isolated particles.}\label{fig:surfaceError}
 \end{figure}

Consider the planar code shown in Fig. \ref{fig:surfaceError}.
Let us say a bit flip error occurs on qubit number $1$ and the wave function of the system changes to
$X_1 \ket{\psi}$. Now, when we measure the stabilizers, assuming a perfect measurement, all syndromes
would be $+1$ except for the measurement outcomes of $Z$ stabilizers marked by blue circles in
Fig. \ref{fig:surfaceError}, which will be $-1$. Thus a single bit flip error creates two adjacent $m$ particles. If instead of a bit flip,
a phase flip error had happened, then it would be the $X$ stabilizers marked by red triangles adjacent
to qubit $1$ that would give different output, giving rise to a pair of adjacent $e$ particles.

An arbitrary single qubit error on the qubit number $1$ would change the wave function of the system to:
\begin{align}
  e^{i \theta \bm n \cdot \bm \sigma_1}\ket{\psi}=&\cos(\theta)\ket{\psi}+i\sin(\theta)n_x X_1\ket{\psi}\nonumber \\
                                                  &+i\sin(\theta)n_y Y_1\ket{\psi}+i\sin(\theta)n_z Z_1\ket{\psi}.\label{equ:generalerror}
\end{align}
The four terms on the right hand side of Eq. \ref{equ:generalerror} have different error syndromes and after one round
of measurement, the system will collapse to one of them. So, from the standpoint of error correction, any
single qubit error reduces to a bit flip or phase flip error, or a combination of the two.

If instead of a single qubit error, many adjacent qubits flip at the same time, only the syndromes at the end of the flipped string
would give different values. Fig. \ref{fig:surfaceError} illustrates an example. An error string, such as $E = \prod_{i \in s} X_i$ for
some string $i$, creates an $m$ particle at each end of the string. Similarly an error string consisting of Pauli-$Z$ errors
creates a pair of $e$ particles at its ends. However if an error string ends on an appropriate boundary, such as a $Z$ error string that
ends on an $e$ boundary, then only a single $e$ particle is created, and therefore only one stabilizer, at the endpoint of the string in the
bulk, is violated.

Therefore, when an error occurs in the form of some strings, the only information we get from syndrome measurements is the
position of the $e$ and $m$ particles. However, given a set of syndrome measurements (locations of $e$ and $m$ particles),
the error string that can create it is non-unique; many different errors can result in the same configuration of particles.
The minimum weight perfect-matching method \cite{edmonds1965a,edmonds1965b} can be used to track back the most likely
error strings from the error syndromes. The method finds the set of shortest possible strings that connect a given set of particles. Since longer error
strings occur with lesser probability, this algorithm finds the most probable error configuration consistent with the measured syndromes.
A logical error occurs when the error string inferred from the minimum weight perfect matching algorithm differs from the
correct error string by a non-contractible string.  On the other hand, if the inferred error strings are always related to the true
error strings by a contractible loop, then that means that we have successfully tracked all of the errors in the software and
can compensate for them accordingly. Other variants of the matching algorithm can be used to improve the probability of guessing the true error configuration\cite{heim2016optimal,baireuther2018neural}.

Since the standard minimum weight matching algorithm runs in polynomial time in system size $l$, for large patches of
surface code other methods like renormalization-group decoders with $\mathcal{O}(\log l)$ run time could become favourable \cite{duclos2010fast,duclos2013fault}. Having enough classical resources, one can also solve the minimum weight matching problem in constant time using parallel computing \cite{fowler2013minimum}.

The probability of a logical error $p_{\text{fail}}$ clearly depends on the underlying error model.
For uncorrelated single qubit errors, numerical and analytical studies suggest an exponential
suppression of $p_\text{fail}$ with increasing code distance\cite{dennis2002,wang2003confinement,Fowler2012proof,Fowler2013analytic,watson2014logical}.
The rate of exponential decay depends on the physical error probability. Specifically, for fixed $d$ and small
probability of physical errors $p$, $p_\text{fail}$ is best described by $A(d)(p/p_\text{th})^{d/2}$ where
$p_\text{th}$ is called the accuracy threshold \cite{Fowler2012proof,Fowler2013analytic,wang2003confinement,fowler2012topological}.
The same form applies for other variants of the surface code but with different values for $p_\text{th}$.

So far we have assumed that the measurement process is perfect. But one also needs to consider the errors that occur
in the measurement process. Measurement errors can be addressed by repeating the measurements many times to distinguish
the measurement errors from other errors. By repeating the measurement many times, we get a three dimensional map for
the position of quasiparticles: two dimensions are used to record the error syndromes in space for each round of measurement
and the third dimension is the discrete time. Now, we use the minimal weight perfect-matching algorithm to connect
the quasiparticles in this three dimensional lattice together, allowing for the strings to have time segments as
well as spatial ones\cite{fowler2012,fowler2012topological}.

The number of measurement histories that are used for error correction depends on the code distance and the
probability of measurement errors. For equal error probability in measurement and storage,
$\mathcal{O}(d)$ rounds of previous error syndromes are used to correct the code where $d$ is the code distance\cite{raussendorf2007,fowler2012towards}.

\subsubsection{Measuring string operators in planar codes}\label{sec:stringmeasurement}

Here we will discuss how to fault-tolerantly measure the string operators associated with $\bar X$ and $\bar Z$. These methods
can also be used for initializing logical qubits in the $\bar X$ or $\bar Z$ basis.

We note that one method to measure $\bar X$ and $\bar Z$ is to measure all physical qubits in the $X$ or $Z$ basis
in order to measure the corresponding logical operator. However since this method is destructive it cannot be used
when there are more than one logical qubits encoded in a patch. In contrast, the string measurement that we review below can be
applied to more general encoding schemes as well.

\begin{figure*}
\vspace{1cm}
\centerline{\includegraphics[width=0.7\textwidth]{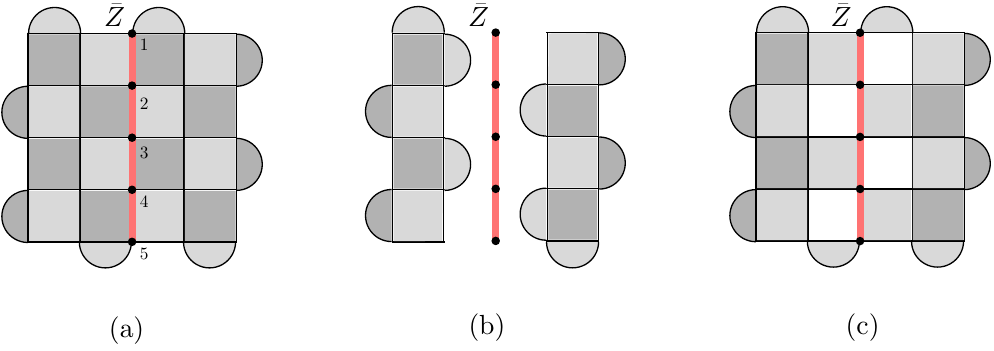}}
\caption{String initialization/measurement method. For initializing the code in an eigenstate of the string operator $\bar Z$ shown in a),
one can turn off the $X$ stabilizers adjacent to the string and change $Z$ stabilizers next to the string to detach the string and
form the code shown in b). Then measure the qubits on the string individually in the $Z$ basis as well as the modified stabilizers
for the sake of error correction, which yields the value of $\bar{Z}$. In this step the code effectively looks like (c). After
correcting errors, we initialize physical qubits in the $\ket{0}$ state and turn back on all stabilizers in their original form.
After $d$ rounds of syndrome measurement and correcting errors, the code is initialized in the $\ket{\bar 0}$ state.}\label{fig:surfaceStringMeasurement}
\end{figure*}

Suppose for example that we wish to measure the string operator $Z_1 Z_2 Z_3 Z_4 Z_5$, shown in Fig. \ref{fig:surfaceStringMeasurement}a. We proceed as follows:
\begin{enumerate}
  \item
  We turn off every $X$ stabilizer that shares a qubit with the string operator $\bar Z$. We also remove
every qubit present in $\bar Z$ from all $Z$ stabilizers adjacent to it, thus changing the $4$ qubit $Z$
stabilizers adjacent to the string operator to a pair of $2$ qubit $Z$ stabilizers. After making these
changes, the code would look like Fig. \ref{fig:surfaceStringMeasurement}b. Note that practically
we have created a new $e$ edge along $\bar Z$.
  \item
  We measure qubits $1$-$5$ individually in the $Z$ basis, in addition to performing the stabilizer measurements. We do $d$ rounds of  measurements to make the the procedure fault tolerant. Using the value of the individual qubit measurements, along with the measurement outcome of the modified stabilizers, we can recover the value of the original $Z$ stabilizers as well. This allows us to track the errors from before the measurement process began.
   \item
 Finally, we turn on all stabilizers and change all the modified stabilizer operators back to their original form.
We need to do $d$ rounds of stabilizer measurements to establish stabilizer values and redefine the code space accordingly.
\end{enumerate}
The measurement of $\bar Z$ is obtained by multiplying the measured values for the individual $Z_i$ measurements along the string.
To make the measurement fault tolerant, it is important to correct any bit flip errors on the $\bar Z$ string before using
individual measurement outcomes in step 2, and to perform the measurement $d$ times to protect against measurement errors. It is worth noting that one can also measure a ribbon of qubits with thickness $d$ once, instead of measuring a string $d$ times. However, to avoid decreasing the code distance, ribbon measurement requires using larger code patches.

Note that phase flip errors that occur on qubits $1$-$5$ will not change the measurement outcome of $\bar Z$ operator.

Measurement in the $\bar X$ basis can be done by following similar steps. However, importantly, measurement of $\bar Y$ cannot
be done in this encoding without introducing additional ingredients, as we describe later.

\subsubsection{Quantum computing with planar codes}

In order to implement universal fault-tolerant quantum computation, we need to implement a universal gate set fault-tolernatly.
For the surface code, a natural choice is the Clifford group, together with the $T$ gate, which is the $\pi/8$ single-qubit phase gate.
Here we will briefly review the proposals for implementing logical Clifford gates in the encoding described above. The $T$ gate is
then implemented fault-tolerantly using magic state distillation.

The Clifford group is generated by the single-qubit Clifford phase gate, $\bar S = \left(\begin{matrix} 1 & 0 \\ 0 & i \end{matrix} \right)$,
$\bar H = \frac{1}{\sqrt{2}}\left(\begin{matrix} 1 & 1 \\ 1 &-1 \end{matrix} \right)$, and the two-qubit CNOT gate.

Note that logical $\bar Z = {\bar S}^2$ is easy to implement, as one can implement it transversally by applying the single-qubit
$Z$ gates on physical qubits along the $\bar Z$ string. $\bar X = \bar H \bar Z \bar H$ can be applied similarly.

The logical Hadamard gate, $\bar H$, is not as straightforward as $\bar X$ and $\bar Z$. Although applying the Hadamard gate
transversally to each individual physical qubit does exchange eigenstates of $\bar X$ and $\bar Z$, it will also change
the boundary conditions, as an $e$ boundary is converted to an $m$ boundary, and vice versa. Therefore,
the transversal Hadamard operation does not yield the original code, but rather yields a $\pi/2$ rotated version of it.
One then needs to correct the orientation by code deformation\cite{dennis2002,bombin2009quantum,horsman2012}.
Code deformation changes the shape of a surface code geometrically by adding physical qubits to the lattice or removing some from it.
Adding and removing here does not mean physical changes to the underlying lattice, but it refers to turning on some stabilizers to include
some idle physical qubits or turning off some stabilizers to exclude some physical qubits from the code.
These additional idle physical qubits add to the spatial overhead required for implementing $\bar H$.

The Clifford phase gate, $\bar S$ is more complicated in current proposals for the planar code. All current proposals
for implementing $\bar S$ in planar encoding described above, require state injection and state distillation. There are proposed hybrid schemes\cite{brown2017} that avoid state distillation for $\bar S$ gate which we will mention shortly.  In
the distillation protocol discussed in Ref. \onlinecite{fowler2012}, a single round of state distillation takes 7 copies
with error probability $p$ and returns one copy  with error probability $7p^3 \ll p$. $k$ rounds
of distillation requires $7^k$ logical qubits. Therefore the spatial overhead grows exponentially in the
number of distillation rounds (See Appendix \ref{distapx}). The overhead of performing the Clifford phase gate is thus extremely high. It is worth noting
that a modified version of the planar code makes it possible to keep track of single qubit Clifford gates
including $\bar S$ at the classical level, thus eliminating the need for state distillation by avoiding direct
implementation of the $\bar S$ gate\cite{litinski2017}.

The two-qubit logical CNOT gate has been proposed to be implemented as follows. One method is to apply CNOT transversally
between every physical qubit in one plane and the corresponding qubit in the other.\cite{dennis2002} However this operation is non-local
if we limit ourselves to a single-layer two-dimensional layout, and thus will not be further considered.

\begin{figure}[!h]
\centerline{\includegraphics[width=0.4\textwidth]{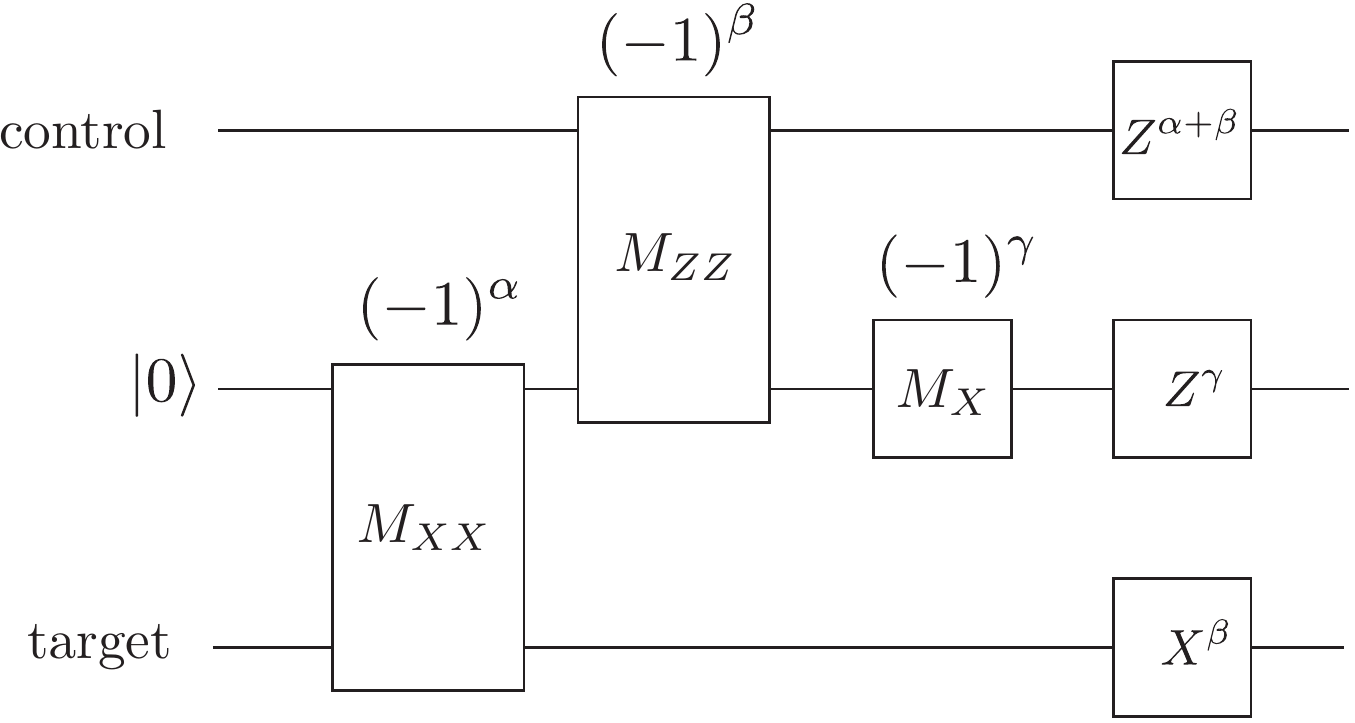}}
\caption{Quantum circuit for CNOT. The number above each measurement represents the outcome of that measurement.}\label{fig:CNOTgate}
\end{figure}

A method for implementing CNOT using local interactions in planar codes uses a method referred to as lattice surgery.\cite{horsman2012}
This method utilizes an extra logical ancilla qubit together with the circuit shown in Fig. \ref{fig:CNOTgate}.\cite{horsman2012}
$M_O$ in the circuit indicates measurement of operator $O$.

We have already explained how to perform the $M_X$ and $M_Z$ measurements. What remains is to explain
how to perform the joint measurement such as $M_{XX}$ and $M_{ZZ}$ in planar codes.
It is important to note that measuring $\bar Z_1$ and $\bar Z_2$ separately and then multiplying the result is
not equivalent to a $\bar Z_1 \bar Z_2$ measurement, as the former will project the code into a smaller subspace than intended.

\begin{figure}[h]
\centerline{\includegraphics[width=0.3\textwidth]{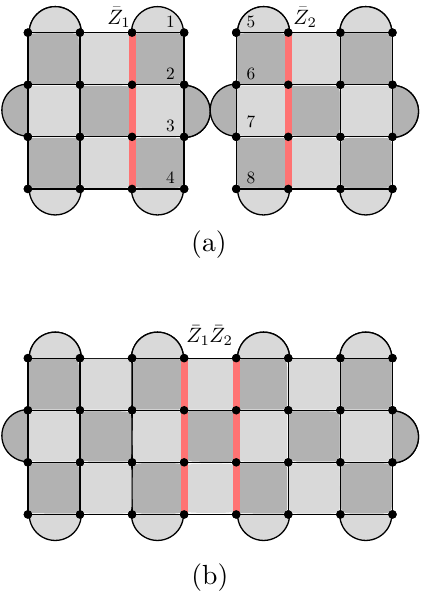}}
\caption{a) Two disjoint planar codes with $\bar Z_i$ operators shown. b) To measure the parity operator $\bar Z_1 \bar Z_2$, we turn on all $Z$ stabilizers that are between two patches. We also combine two $2$-qubit $X$ stabilizers at the boundary into one full stabilizer. After $d$ rounds of syndrome measurement and error correction we can find the value of $\bar Z_1 \bar Z_2$ by multiplying the newly measured $Z$ stabilizers.  }\label{fig:merge}
\end{figure}

Consider two planar codes next to each other, as in Fig. \ref{fig:merge}a. Note that the neighboring boundaries are
both $m$ boundaries. To measure the two body operator $\bar Z_1 \bar Z_2$ we use the following steps:
\begin{enumerate}
  \item
 We stop measuring the $X_2 X_3$ and $X_6 X_7$ stabilizers and start to measure the combined $X_2 X_3 X_6 X_7$ stabilizer.
At the same time, we start measuring two new $Z$ stabilizers $Z_1 Z_2 Z_6 Z_5$ and $Z_3 Z_4 Z_8 Z_7$. This modification
effectively merges the two patches together and the code will look like Fig. \ref{fig:merge}b.
  \item
  We wait for $d$ rounds of stabilizer measurements to establish the values of newly added $Z$ stabilizers.
  \item
  We read the value of $\bar Z_1 \bar Z_2$ by multiplying the measurement outcomes of newly added $Z$ stabilizers. After that we stop measuring all three shared stabilizers and turn back on $X_2 X_3$ and $X_6 X_7$ stabilizers to detach the codes again.
\end{enumerate}

If the patches are oriented in such a way that $e$ boundaries are next to each other, we can
measure the $\bar X_1 \bar X_2$ operator by turning on the shared $X$ stabilizers. The procedure is similar to the $M_{ZZ}$ measurement.

\begin{figure}
\centerline{\includegraphics[width=0.3\textwidth]{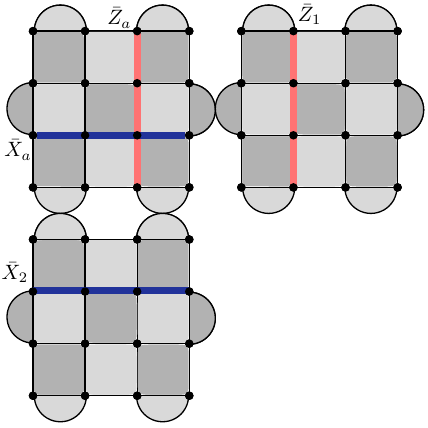}}
\caption{Planar code layout to do the CNOT. The patch in the corner encodes the ancilla qubit, the one in the right is the control qubit and the bottom one is the target qubit.  }\label{fig:ZZXX}
\end{figure}

Using the joint measurements, the quantum circuit shown in Fig. \ref{fig:CNOTgate} can be implemented by using the configuration
shown in Fig. \ref{fig:ZZXX}. The logical qubit in the corner is the ancilla qubit, the bottom patch encodes the target qubit and the
other one is the control qubit. Note that the patches are oriented in such a way to make the joint measurements in
Fig. \ref{fig:CNOTgate} possible. In the last step, we need to apply single qubit gates based on the outcome of previous
measurements as shown in Fig. \ref{fig:CNOTgate}.

\subsection{Hole encoding}

\begin{figure}
\centerline{\includegraphics[width=0.4\textwidth]{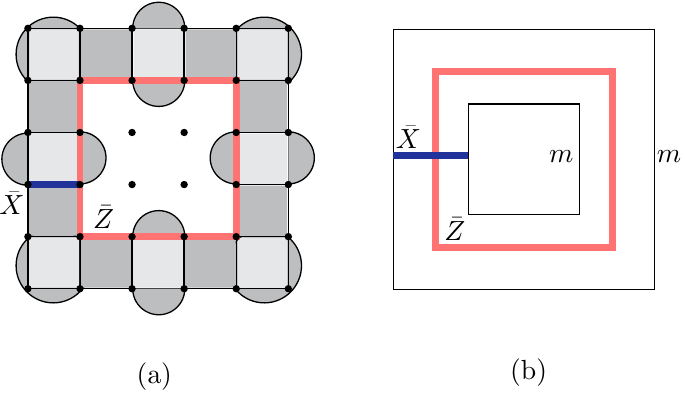}}
\caption{a) Surface code with hole defect in the bulk. Note that both the outer boundary and the hole's boundary are $m$ boundaries.
b) Schematic diagram for the lattice structure shown in (a). }\label{fig:SurfaceHole}
\end{figure}

If we start with a planar code and remove some qubits from the bulk, we obtain a hole defect. Fig. \ref{fig:SurfaceHole}a shows a
hole defect that is created by turning off nine stabilizers. Although the qubits inside the hole are completely detached
from the code, they are needed for moving the hole. Each hole introduces new edges and like the outer edges, the
boundary of a hole can be either an $e$ edge or $m$ edge. In principle a hole can have mixed boundary conditions, but
usually uniform boundaries are used.

Depending on the boundary type, $e$ or $m$ particles can condense on a hole boundary. This in turn allows
one to use hole defects to make new logical operators and thus new logical qubits. The hole defect in
Fig. \ref{fig:SurfaceHole}, for example, encodes one logical qubit.

\begin{figure}
\centerline{\includegraphics[width=0.3\textwidth]{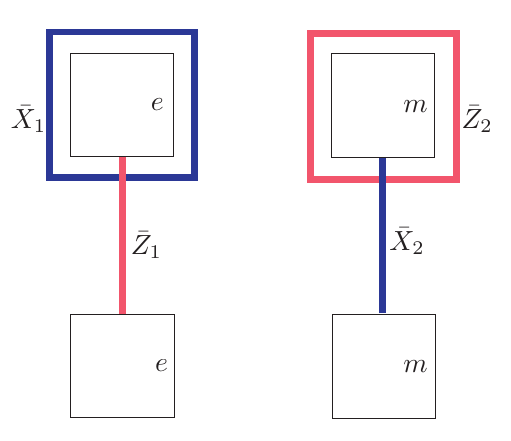}}
\caption{Logical qubits encoded in pairs of hole defects alongside the corresponding logical operators. The outer boundary of the code can be far away and is not shown here. }\label{fig:2holes}
\end{figure}

In general, $n$ holes of the same boundary type define a $2^{n-1}$ dimensional code subspace.
The proposal described in Ref. \onlinecite{fowler2012}, however, uses a sparse encoding, where each logical qubit is
encoded using two holes, as shown in Fig. \ref{fig:2holes}. A logical qubit that is defined using
a pair of $e$ boundaries is called a $X$-cut qubit (Fig. \ref{fig:2holes} left). Likewise,
$Z$-cut qubit refers to a logical qubit encoded in a pair of $m$ boundaries (Fig. \ref{fig:2holes} right).

The sparse encoding allows for implementation of logical gates as described below.
Our joint measurement technique, described in the subsequent sections,
allows Clifford operations to be implemented using denser encodings, and thus may offer advantages in overhead.

\subsubsection{Quantum computing with hole defects}

The measurement and application of the logical $\bar X$ and $\bar Z$ operators proceeds analogously to the
case of the planar encoding. The single qubit Clifford phase gate, $\bar S$, is also proposed to be implemented
using state injection and state distillation, as in the case of the planar encoding. Similar to the planar code, one can circumvent distillation by using the hybrid schemes \cite{brown2017} which we will mention shortly.

The single qubit Hadamard gate $\bar H$ is performed through a series of code deformations\cite{fowler2009high, fowler2012}, as follows.
Assume we have a pair of $e$ holes encoding our logical qubit. Since, unlike the planar code, there is generally more than one logical qubit
encoded in a patch, first we isolate the target logical qubit from the rest by measuring a Pauli $X$ string which encircles the hole pair. As
was explained in Sec. \ref{sec:stringmeasurement}, this would create an $m$ boundary around the two $e$ holes. By expanding
the holes one can turn them into $e$ boundaries of the isolated patch, converting the logical qubit to a planar encoding using boundary defects.
The Hadamard gate is then applied as it is in the planar code, described above, and then finally the logical qubit is converted back to
the hole encoding and merged into the rest of the code.

The logical CNOT operation is quite different in the hole encoding as compared with the planar encoding.
If we have a $Z$-cut qubit and a $X$-cut qubit, one can show that moving a hole of one qubit around a hole of the other,
will perform CNOT between the two\cite{raussendorf2006,fowler2009high, fowler2012}. This process is called hole braiding.
However performing CNOT between two qubits with the same type of holes is more complicated, because braiding
two holes of the same boundary type is a trivial operation in the code subspace. Instead, in this case one
needs extra logical ancilla qubits encoded using holes with the other type of boundary. One can then implement the CNOT gate between two hole defects of the same type
through a series of hole braidings and measurements \cite{fowler2012}. Therefore, to perform a CNOT
on two logical qubits requires a total of six holes, if the two logical qubits are both $X$- or $Z$- cut qubits.

\subsection{Dislocation encoding}\label{sec:dislocationencoding}
\begin{figure}
\centerline{\includegraphics[width=0.5\textwidth]{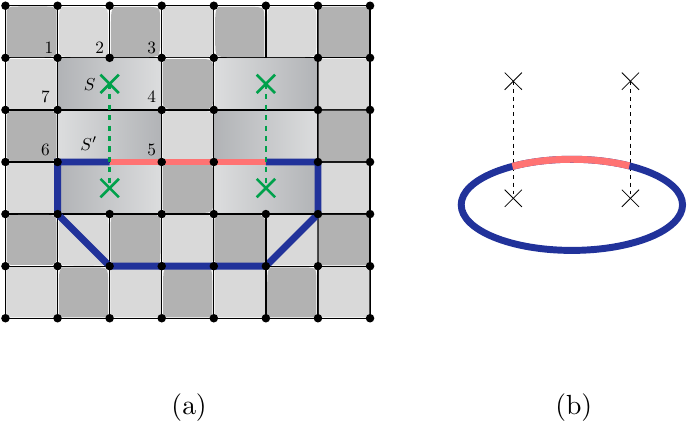}}
\caption{(Color online) (a) A surface code with four twist defects marked by green crosses. The boundary of the surface code is not
important and hence has not been shown. Green dashed lines are called dislocation lines. Every twist defect should
be connected to another twist defect by a dislocation line. To make a pair of twist defects, we start with a perfect lattice
and draw the dislocation line. Then we remove every qubit that lies on the dislocation line. Finally, we combine every
two stabilizers that share an edge over the dislocation line into one stabilizer which is given by the product of
original stabilizers. Clearly, the contribution of removed qubits have to be omitted. The combined stabilizer is
 represented by plaquettes with color gradient since they are neither $X$ nor $Z$ stabilizer, but have both operators.
It is also possible to create dislocations without removing qubits by having the dislocation lines parallel
to the Burgers vectors of the dislocations (not shown here)\cite{bombin2010,barkeshli2012a}.
(b) Schematic diagram of the surface code shown in (a).}\label{fig:surfacetwist}
\end{figure}

Making holes inside the bulk is not the only way to introduce non-trivial closed loops in surface codes.
Twist defects can also be used to induce topological degeneracies and thus to encode logical qubits.
Twist defects have been studied from a number of points of view, using topological field theory
(see Sec. V of Ref.\onlinecite{barkeshli2010} and Ref.\onlinecite{barkeshli2012a,barkeshli2013genon,barkeshli2013defect2,teo2013,barkeshli2014SDG}),
chiral Luttinger liquid theory\cite{barkeshli2012a,barkeshli2013genon,barkeshli2013defect2,clarke2013,cheng2012,lindner2012,alicea2016review},
and in lattice models for topological order\cite{bombin2010,kitaev2012,you2012,you2013}.

Fig. \ref{fig:surfacetwist}a illustrates a surface code with four twist defects in the bulk which are
marked by green crosses. As is clear from Fig. \ref{fig:surfacetwist}a, the physical qubits on
the dislocation lines are removed from the lattice and all pairs of stabilizers that share an edge over the
dislocation line are combined into one.\footnote{Fig. \ref{fig:surfacetwist} illustrates dislocation lines that run perpendicularly to their Burgers vectors.
It is also possible to consider a lattice geometry with dislocation lines parallel to the Burgers vectors, so that qubits do not need to be removed to create
a dislocation.} The stabilizers located on a twist defect involve five physical qubits,
and one of the qubits should be measured in the $Y$ basis. For example, the stabilizer $S$ in Fig. \ref{fig:surfacetwist}a is defined as:
\begin{equation}
  S=X_1 Y_2 Z_3 Z_4 X_7,
\end{equation}
and the stabilizer corresponding to the plaquette just below that is:
\begin{equation}
  S'=Z_7  X_4 X_5 Z_6.
\end{equation}

As with hole defects, non-trivial string operators encircling twist defects can be used to define logical qubits. But an
important property of dislocation lines is that whenever a string operator passes through them, it changes its type;
a Pauli-$Z$ string would change to a Pauli-$X$ string and vice versa. As a result, a closed string operator
needs to encircle at least two twist defects. A non-trivial closed string operator is shown as an example at the
bottom of Fig. \ref{fig:surfacetwist}a and it includes both Pauli-$X$ and Pauli-$Z$ operators:
\begin{equation}
\bar Z=\prod_{i\in \text{light red part}}Z_i\prod_{j\in \text{dark blue part}}X_j.
\end{equation}
One can easily verify that this operator commutes with every stabilizer but is not a product of stabilizers itself.
Fig. \ref{fig:surfacetwist}b shows the schematic diagram of the code shown in Fig. \ref{fig:surfacetwist}a.

In general, $n$ pairs of twist defects gives rise to $2^{n-1}$ states. In a dense encoding, therefore, there would be one logical qubit
for every pair of twist defects (not counting the first pair). Alternatively, sparser encodings are also possible, using three or four twist defects
to encode one logical qubit.

\begin{figure}
\centerline{\includegraphics[width=0.5\textwidth]{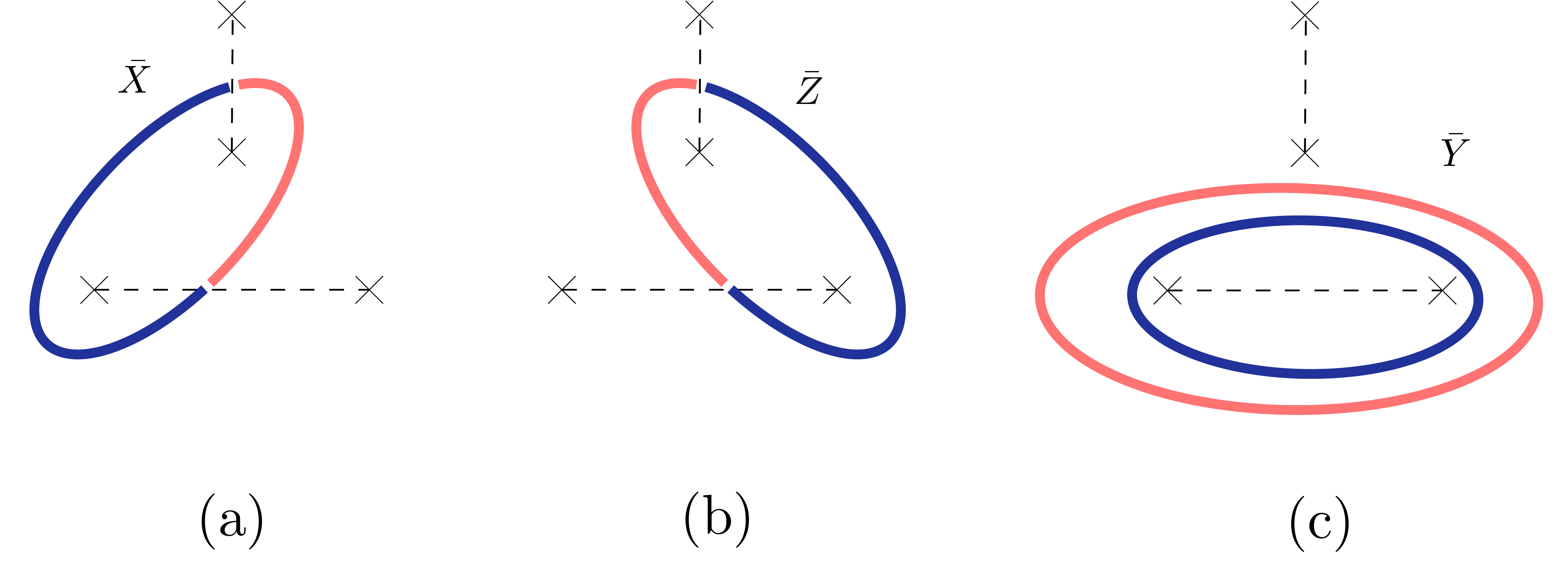}}
\caption{(Color online) Logical operators related to a qubit encoded in three twists. (a) String operator corresponding to $\bar X$. It is a
mixed Pauli string, with red (light gray) parts corresponding to Pauli $X$ and blue (dark gray) parts corresponding to Pauli $Z$. Note that
whenever the string passes through a dislocation line, it changes color. An equivalent string operator for $\bar X$
would be the one with red (light gray) and blue (dark gray) interchanged. All that matters is the defects a string enclose.
(b) String operator corresponding to $\bar Z$. (c) Two equivalent string operators representing $\bar Y$.}\label{fig:twistlogical}
\end{figure}

\begin{figure}
\centerline{\includegraphics[width=0.2\textwidth]{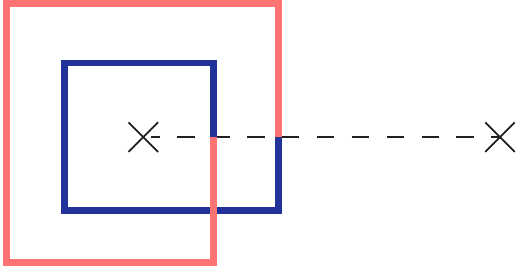}}
\caption{The string that encircles a single twist defect twice and encloses itself can be written
as a product of stabilizers and hence acts as the identity operator in the code subspace.}\label{fig:trivialDoubleLoop}
\end{figure}

A key feature that distinguishes the dislocation code from planar and hole encodings is that the logical $\bar Y$ operator is also
given by a simple Pauli string. Therefore the $\bar Y$ operator can be straightforwardly measured fault-tolerantly
using the same methods for measuring $\bar X$ and $\bar Z$ operators in the planar and hole encodings. Alternatively,
logical qubits can be initialized in the $\bar Y$ basis straightforwardly. Fig. \ref{fig:twistlogical} shows the logical
$\bar X$, $\bar Y$ and $\bar Z$ operators for a logical qubit encoded in three twist defects. It can be shown that any
two loop operators that encircle the same set of twist defects are equal to each other up to multiplication by some set
of stabilizers and hence represent the same logical operator. For example, both Pauli-$X$ and Pauli-$Z$ strings shown in
Fig. \ref{fig:twistlogical}c are equal to $\bar Y$. To prove their equivalence one can use the fact that if a string goes around a
single twist twice and closes itself, it acts as the identity on the code subspace (Fig.~\ref{fig:trivialDoubleLoop}).

\subsubsection{Initialization and measurement}

Again, the string initialization and measurement that was described in section \ref{sec:stringmeasurement} can be
used here too. It is notable that by using string initialization, one can prepare the $\ket{Y}$ state without using state injection and state distillation procedures.

\subsubsection{Quantum Computation by twist defects}

The fact that one can measure a logical qubit in the $\bar Y$ basis as well as $\bar X$ and $\bar Z$, allows one to
ignore every single-qubit Clifford gate until there is a measurement, and then modify the measurement according to the
awaiting gates\cite{hastings2014}. For example if we want to apply an $S$ gate on a logical qubit and measure
it in the $X$ basis, we can ignore the first gate and instead do the measurement in the $S^{\dagger} X S=Y$ basis. This point is explained in more detail in section \ref{sec:classicalTracking}.

We can also easily implement CNOT using the circuit shown in Fig. \ref{fig:CNOTgate}. Joint measurements in
dislocation codes are not really different from single qubit measurements. Suppose we want to measure
the $\bar X_1 \bar X_2$ operator where $\bar X_1$ is given by a string encircling twists $1$ and $2$ and
$\bar X_2$ encircles twist defects numbered $3$ and $4$. It is easy to see that $\bar X_1 \bar X_2$ is
given by the simple string that encircles all four twist defects $1$ to $4$. We will explain this point further
in Sec. \ref{sec:jointsurface}, since joint measurements and twist defects lie at the heart of our method.

\subsection{Triangular code}

\begin{figure}
\centerline{\includegraphics[width=0.5\textwidth]{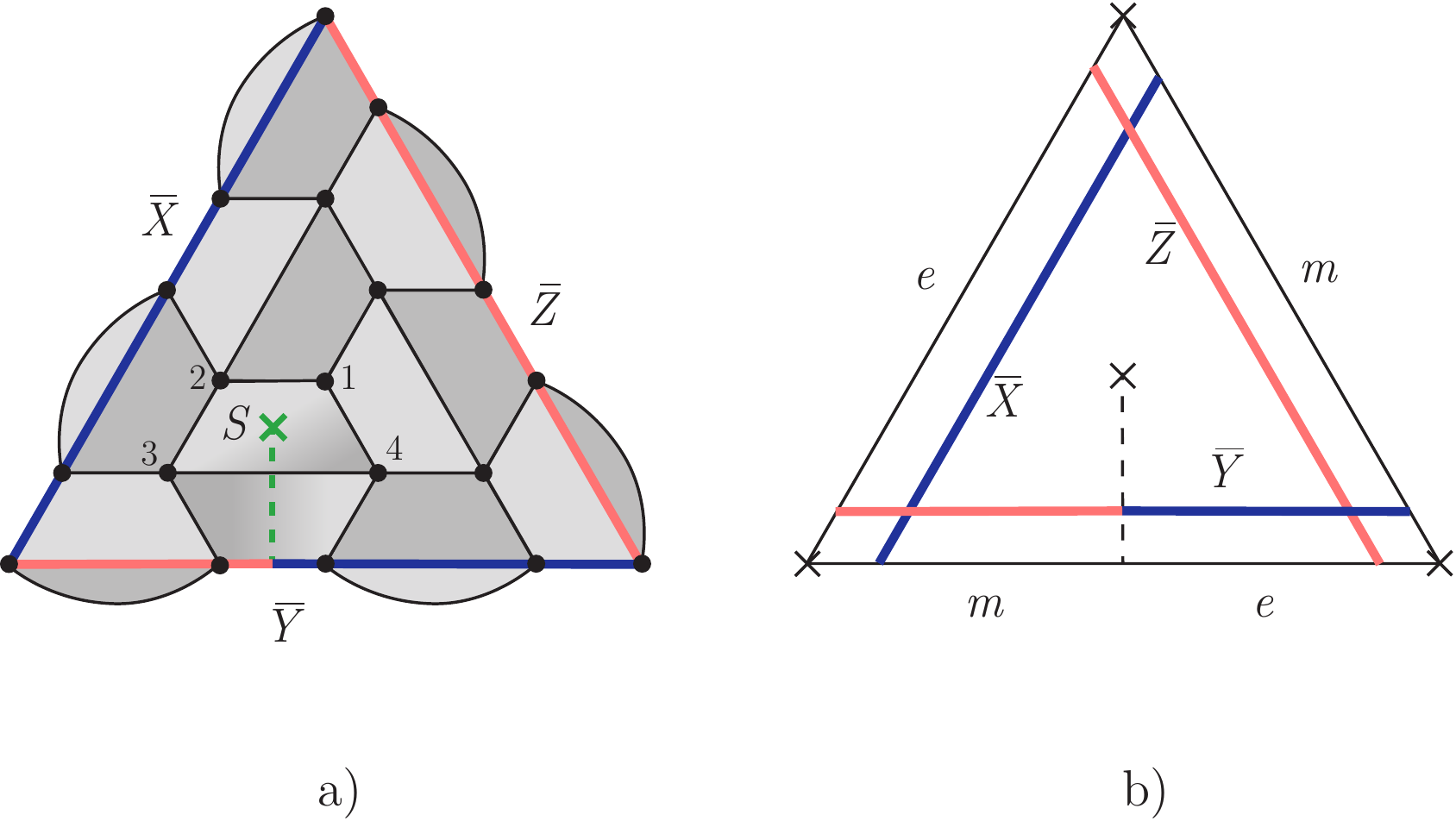}}
\caption{a) A distance $d=5$ triangular code. b) Schematic diagram of the triangular code shown in (a).}\label{fig:tcode}
\end{figure}

Another important variant of the surface code is called the triangular code\cite{yoder2017}. Fig. \ref{fig:tcode}
shows a triangular code patch which encodes one logical qubit alongside its schematic diagram. The lattice
structure results from flattening a square lattice that resides on three adjacent faces of a cube in three dimensions.
There are three boundary defects and one bulk twist defect. Unlike normal twist defects that we mentioned before,
the stabilizer $S$ at the position of the bulk twist defect involves four qubits and is given by $S=Y_1 Z_2 Z_3 Z_4$.
An important distinguishing feature is that in addition to $\bar X$ and $\bar Z$, the $\bar Y$ operator is also
given by a simple Pauli string that starts and ends over the edges. As a result, the code can be initialized or
be measured in the $\bar Y$ basis easily.
A key advantage of the triangular code is that the Clifford phase gate, $\bar S$, can be implemented without state distillation,
and the general overhead requirements are better than the planar and hole encodings described above.

\subsection{Hybrid schemes}

Hybrid schemes for encoding logical qubits are also possible\cite{delfosse2016,brown2017,litinski2017}.
Mixing two schemes also opens up the possibility of new methods to perform logical operations. For example,
by converting boundary defects to bulk twist defects, braiding them and converting them back to the boundary,
one can implement the $\bar S$ gate in planar codes.\cite{brown2017} Alternatively, logical qubits in different encodings can
be entangled with each other, for example by braiding holes and twist defects.

\section{Measurement-based protocols for Clifford gates}\label{sec:circuits}

Here we explain how one can implement all gates in the Clifford group using circuits based on joint measurements.
We only discuss the quantum circuits corresponding to the logical gates, regardless of the underlying setup
which is used to encode logical qubits. In the subsequent sections, we will show how one can implement these
circuits in surface codes, color codes and hyperbolic codes.

In this section all operators are understood to be logical operators, so we will omit the $\bar{\;}$ notation; all qubits are
understood to be logical qubits.

\subsection{CNOT gate}

We have already mentioned the quantum circuit devised to implement
CNOT using joint measurements (Fig. \ref{fig:CNOTgate}).
It is used in many variants of surface codes as well as color codes to implement the CNOT
gate\cite{horsman2012,hastings2014,landahl2014,litinski2017}.

\subsection{S gate}

\begin{figure}
\centerline{\includegraphics[width=0.4\textwidth]{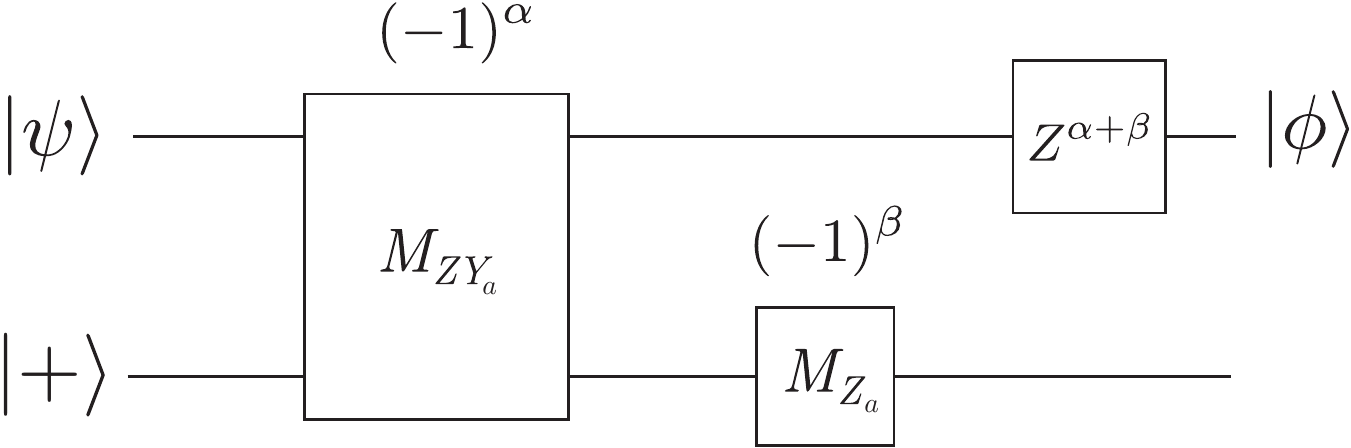}}
\caption{Quantum circuit for S gate. The ancilla is prepared in $\ket{+}=(\ket{0}+\ket{1})/\sqrt{2}$ state. The
joint measurement corresponds to measuring the parity of the operator $Z Y_a $, where the subscript $a$ stand for ancilla qubit
and $Z$ is associated with the data qubit. The outcome of each measurement is written above its box.}\label{fig:Sgate}
\end{figure}

The circuit that is shown in Fig. \ref{fig:Sgate} can be used to implement the $S$ gate. Initially, the ancilla qubit is
prepared in the $\ket{+} $ state, which is the $+1$ eigenstate of the Pauli $X_a$ operator. Next, the two qubit parity
operator $Z Y_a$ is measured, followed by a $Z_a$ measurement. The subscript $a$ is used to distinguish the
operators associated with the ancilla qubit. The Pauli operators associated with the data qubit will have no subscript.

After the second measurement in Fig. \ref{fig:Sgate}, the state of the data qubit is given by:
\begin{equation}\label{equ:1}
\frac{1-i(-1)^{\alpha+\beta}Z}{\sqrt{2}}|\psi\rangle,
\end{equation}
where $(-1)^\alpha$ and $(-1)^\beta$ are the the results of first and second measurements. Note that the $S$ gate can be written as:
\begin{equation}\label{equ:2}
S=e^{i\pi/4}\qty(\frac{1-iZ}{\sqrt{2}}).
\end{equation}
Thus, if the outcome of the two measurements have the same sign, the state after the second measurement is, up to an overall phase,
$S\ket{\psi}$, and therefore the $S$ gate has been implemented.

On the other hand, if the results of the two measurements are different, the state of the data qubit would be $S^\dagger \ket{\psi}$.
Since $S=Z S^\dagger$, we can recover $S \ket{\psi}$ by applying an additional $Z$ gate to the data qubit.

\subsection{SHS gate}

\begin{figure}[!h]
\centerline{\includegraphics[width=0.4\textwidth]{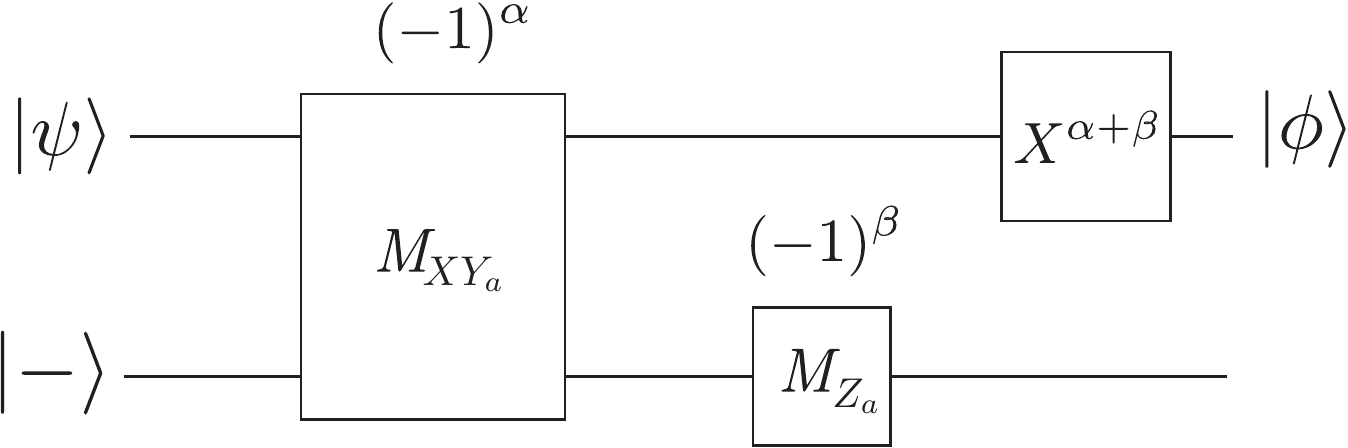}}
\caption{Quantum circuit to implement $SHS$ gate. The ancilla is prepared in the $\ket{-}=(\ket{0}-\ket{1})/\sqrt{2}$ state.
The joint measurement corresponds to measurement of the parity of the operator $X Y_a $, where subscript $a$ stands for ancilla qubit and
$X$ is associated with the data qubit. The outcome of each measurement is written above its box.}\label{fig:Hgate}
\end{figure}

We need another independent gate to fully implement the Clifford group. Usually it is the Hadamard gate $H$, but here we
choose $SHS$ since it has a simpler circuit. The circuit is shown in Fig. \ref{fig:Hgate}. Again, it is easy to check that
after the second measurement, the data qubit corresponds to:
\begin{equation}\label{equ:3}
|\phi\rangle=\frac{1+i(-1)^{\alpha+\beta}X}{\sqrt{2}}|\psi\rangle,
\end{equation}
where $(-1)^\alpha$ and $(-1)^\beta$ are measurement results. Similar to Eqn. \ref{equ:2}, we have:
\begin{equation}\label{equ:4}
SHS=\qty(\frac{1+iX}{\sqrt{2}}).
\end{equation}
So if the results of the two measurements have the same sign, we get the desired state $SHS\ket{\psi}$. Again, if we get
different signs, the data qubit would be in the state $S^\dagger H S^\dagger\ket{\psi}$. We can then recover $SHS \ket{\psi}$
by applying $X$, because $SHS=i X S^\dagger H S^\dagger$. Hence, at the end of the circuit, we get $\ket{\phi}= SHS\ket{\psi}$.

\subsection{Conjugated CNOT circuits}\label{sec:classicalTracking}

Since single qubit Clifford gates just permute the Pauli matrices (up to a sign), like the $H$ gate that exchanges
$X$ and $Z$, it turns out that one can just keep track of them classically instead of actually applying them at
the quantum level. One way to see this is to move all single qubit Clifford gates to the end of the circuit and
then modify the final measurements accordingly. The price to pay is that for a general quantum circuit, we need
to be able to implement CNOT and the $\pi/8$ phase gate $T$ in any Pauli basis. In other words, we need to be able to
implement CNOT and $T$ conjugated by any single-qubit Clifford gate.

\begin{figure}
\centerline{\includegraphics[width=0.5\textwidth]{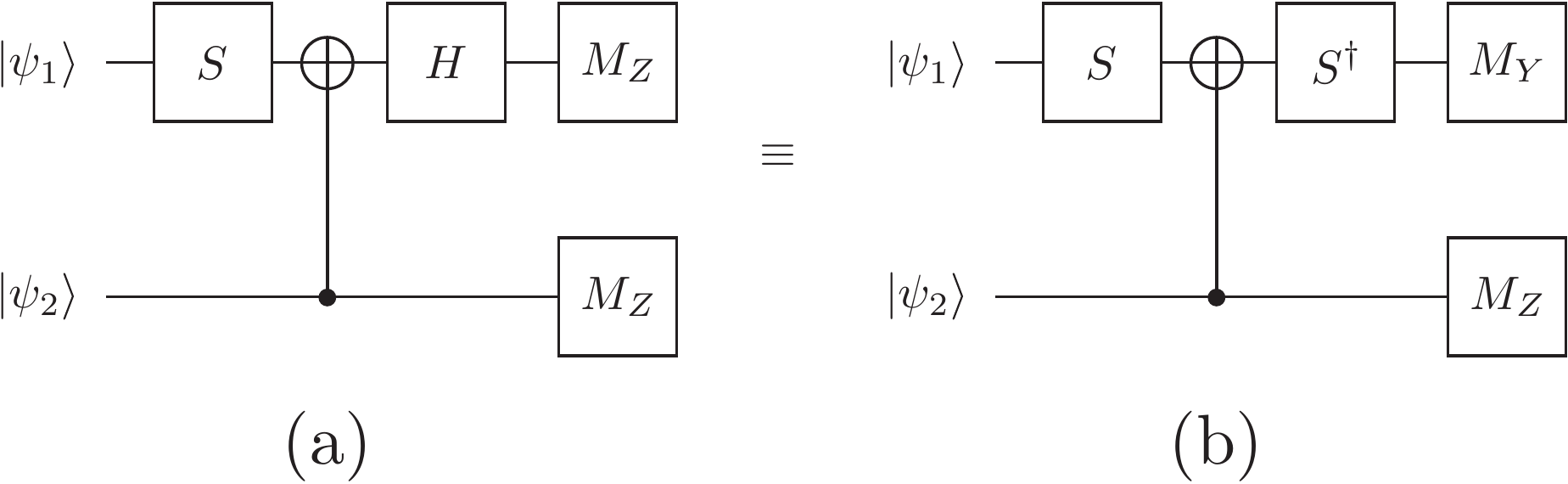}}
\caption{The two quantum circuits in (a) and (b) yield equivalent measurement outcomes, provided the measurements
are done in different bases. }\label{fig:classicalTracking}
\end{figure}

\begin{figure}
\centerline{\includegraphics[width=0.4\textwidth]{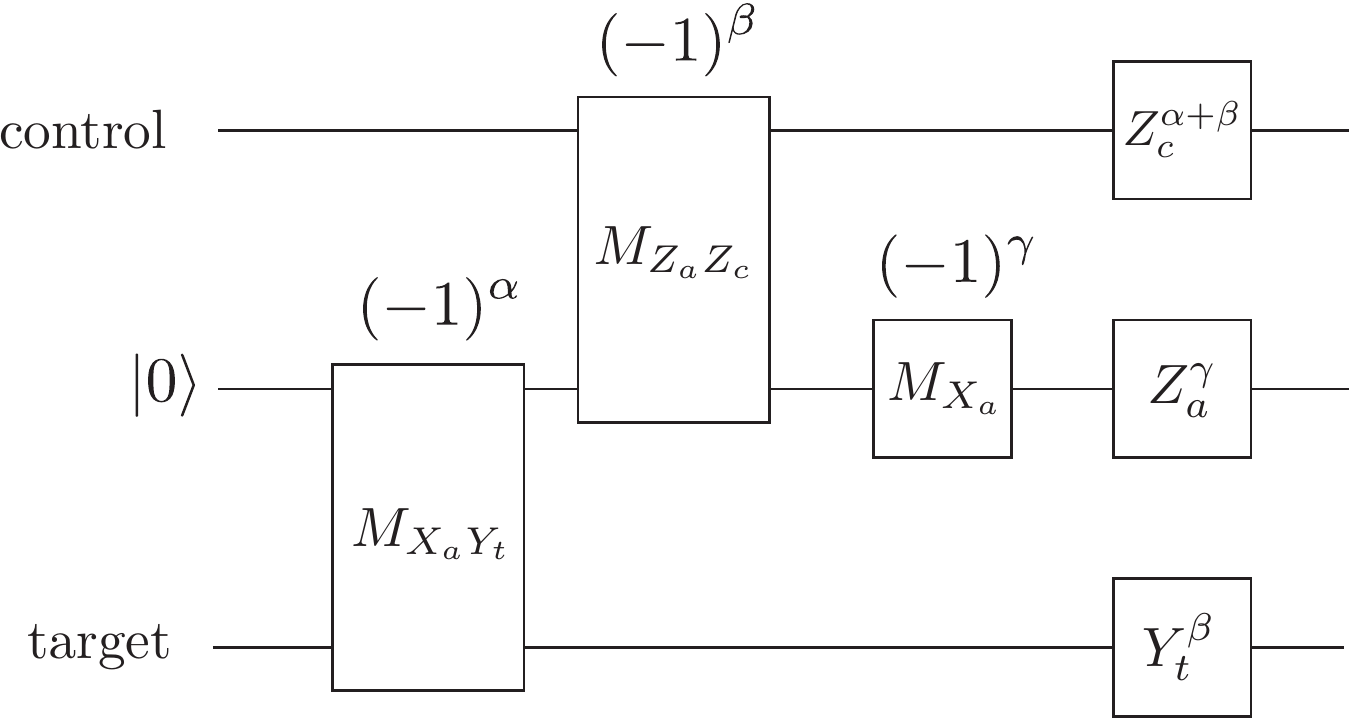}}
\caption{$S_t^\dagger~\text{CNOT}~S_t$ circuit. Here the subscripts $a$, $c$, and $t$ refer to the ancilla, control, and target qubits, respectively. }\label{fig:SCNOTS}
\end{figure}

As an example, consider the quantum circuit shown in Fig.~\ref{fig:classicalTracking}a. We use the
identity  $\text{CNOT}~S=S~(S^\dagger~\text{CNOT}~S)$ to move the $S$ gate across CNOT. Also since
$S^\dagger H^\dagger \Pi_{\pm,Z} HS=\Pi_{\pm,Y}$ , where $\Pi_{\pm,\sigma}$ denote the projection operator
onto the $\pm$ eigenspace of $\sigma$ operator,  we can replace the upper $Z$ measurement at the
end of the circuit with a $Y$ measurement. So the probabilities for each measurement outcome of this quantum
circuit will be equivalent to performing instead $S^\dagger~\text{CNOT}~S$, followed by measurement in a
different basis (Fig.~\ref{fig:classicalTracking}b). The quantum circuit for $S^\dagger~\text{CNOT}~S$
(Fig.~\ref{fig:SCNOTS}) can be derived from the CNOT quantum circuit in Fig.~\ref{fig:CNOTgate}.

\section{Joint measurement scheme in surface codes}
\label{sec:surface}
The measurement circuits for the implementation of the $\bar S$, $\bar S \bar H \bar S$, and $\overline{\text{CNOT}}$ gates discussed above require the ability to perform
joint measurements of operators such as $\bar Z \bar Z_a$, $\bar X \bar X_a$, and $\bar Z \bar Y_a$. Alternatively, if the logical ancilla can be prepared in the $\bar Y$ basis,
then we only need the joint measurements $\bar Z \bar Z_a$ and $\bar X \bar X_a$.

In this section we will discuss two possible procedures to perform such tasks in surface code. First we explain how one can utilize CAT states to perform the required joint measurements. However, as we will discuss shortly, this method is not practical for large code distances and hence, we introduce the second method which utilizes twist defects to carry out the required measurements.
\subsection{Utilizing CAT states with any encoded ancilla}
\label{sec:cat}

\begin{figure}
\centerline{\includegraphics[width=0.4\textwidth]{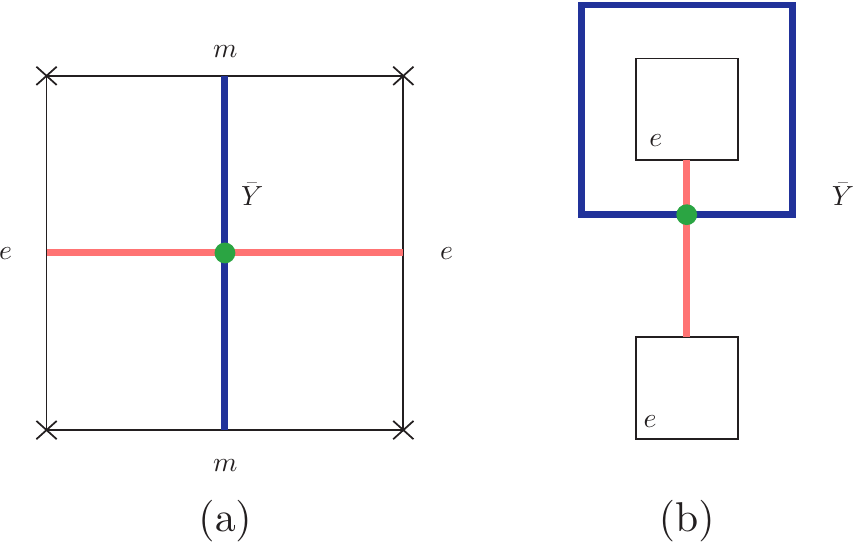}}
\caption{(Color online) a) $\bar Y$ operator for a logical qubit encoded in a planar code. b) $\bar Y$ operator in hole encoding.
The intersection of the red (light gray) and blue (dark gray) strings (marked by the green dot)
corresponds to a Pauli-$Y$ operator.}\label{fig:Ybar}
\end{figure}

In many encoding schemes, such as the purely hole or boundary defect based encodings, measuring the logical $\bar Y$ operator
fault-tolerantly is non-trivial, as the same schemes discussed above for measuring $\bar X$ and $\bar Z$ do not work.
Fig. \ref{fig:Ybar} illustrates the $\bar Y$ operator in the hole and boundary defect based encodings, which has support on a graph, as opposed
to a string. As such, we refer to it as a graph operator. We see, therefore, that $\bar Y$ is a graph operator that contains a
$Y$ acting on at least one physical qubit. If we were to directly measure the individual physical qubit operators along the graph,
then measuring $Y$ would not give us information about the neighboring $X$ and $Z$ stabilizers, and thus the measurement
cannot be made fault-tolerant.

\begin{figure}
\centerline{\includegraphics[width=0.3\textwidth]{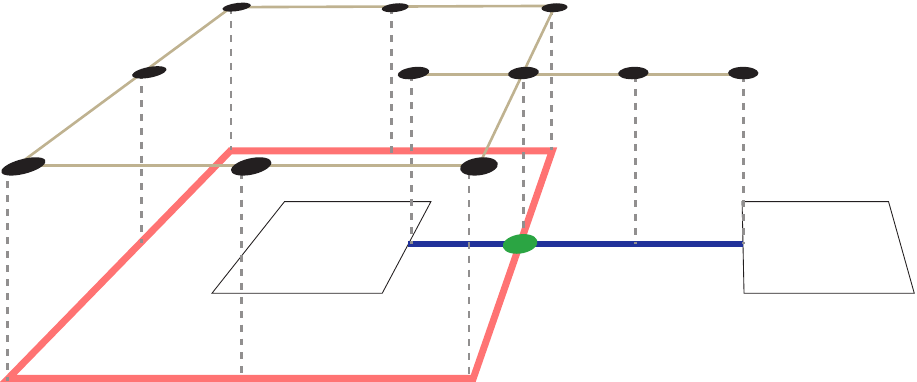}}
\caption{Arrangement of physical ancilla qubits (upper black dots) comprising the CAT state, which is used for fault tolerant measurement of $\bar Y$.}\label{fig:CAT3D}
\end{figure}

\begin{figure}
\centerline{\includegraphics[width=0.5\textwidth]{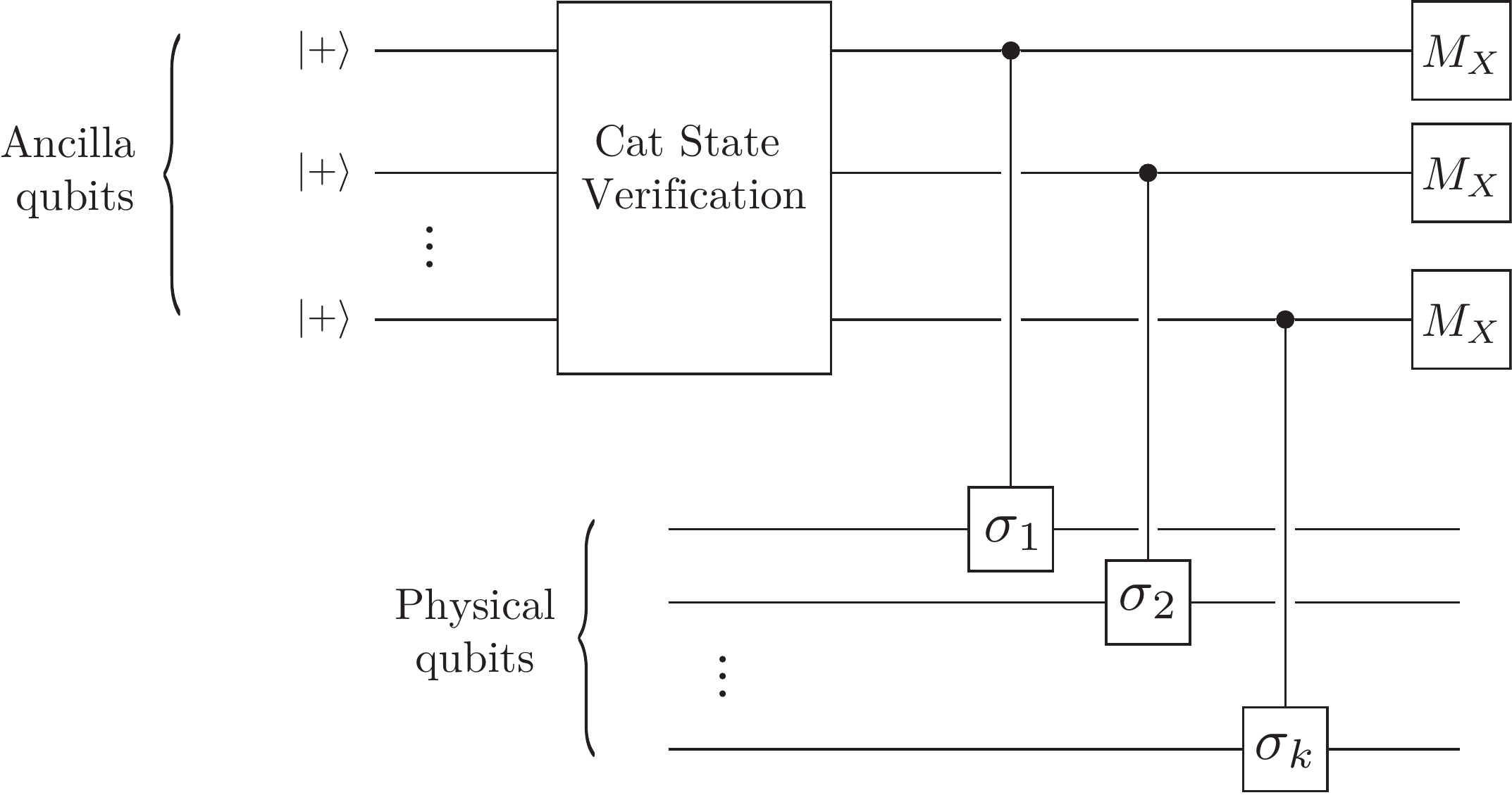}}
\caption{The quantum circuit for preparing a CAT state and measuring a general graph operator of the form $S=\prod_{i=1}^k\sigma_i$
over physical data qubits. The verification part consists of measuring $Z_iZ_{i+1}$ stabilizers $\mathcal{O}(d)$ times and using the
result to correct possible bit flip errors.\cite{brooks2013} This would initialize the ancilla qubits into the CAT state. Then, Controlled-Pauli
gates are applied between each ancilla qubit and the corresponding data qubit. Afterwards, all ancilla qubits are measured in the $X$ basis.
The logical measurement outcome would be equal to the parity of theses individual measurements. The whole measurement procedure
is made fault-tolerant by repeating it $\mathcal{O}(d)$ times to obtain a majority vote.}\label{fig:CAT_circuit}
\end{figure}

However, we can perform the measurement of $\bar Y$ fault-tolerantly with the aid of a CAT state consisting of $k \propto d$ physical
ancilla qubits that run along the graph operator, as shown in Fig. \ref{fig:CAT3D}. As discussed in Ref. \onlinecite{brooks2013}, a CAT state
can be prepared with local operations by preparing each qubit in the CAT state in the $\ket{+}$ eigenstate, and then measuring the products $Z_i Z_{i+1}$
for nearest neighbor qubits. Armed with the CAT state, we can then measure $\bar Y$ using the circuit illustrated in Fig. \ref{fig:CAT_circuit}.
We refer the reader to Refs. \onlinecite{brooks2013,nielsen2002quantum} for a detailed discussion of the fault-tolerant preparation of CAT states with local operations.

\begin{figure}
\centerline{\includegraphics[width=0.5\textwidth]{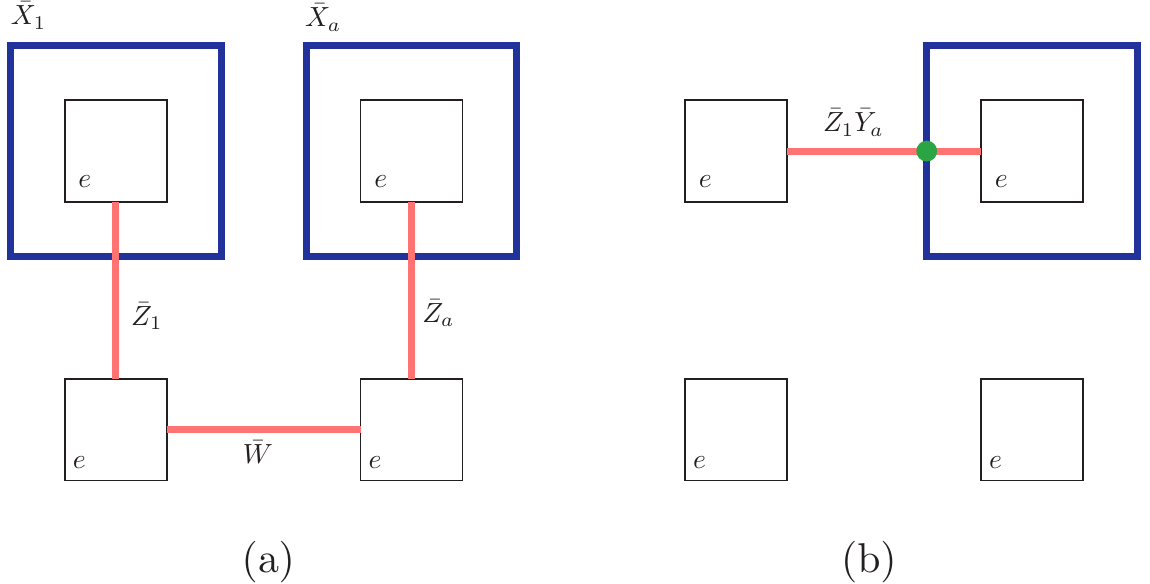}}
\caption{a) A logical data qubit and a logical ancilla qubit encoded in four hole defects. The $\bar W$ string does not encode any logical information and can be initialized to $+1$. This \textit{idle} string would be used to facilitate joint measurements. b) The graph operator corresponding to $\bar Z_1 \bar Y_a$ can be measured fault tolerantly using CAT states.}\label{fig:ZY}
\end{figure}

These CAT states can therefore be utilized to either prepare the logical ancilla in an eigenstate of $Y$
by measuring $\bar Y_a$ or, alternatively, to measure the operator $\bar Z \bar Y_a$, as shown in Fig. \ref{fig:ZY}.

\begin{figure}
\centerline{\includegraphics[width=0.3\textwidth]{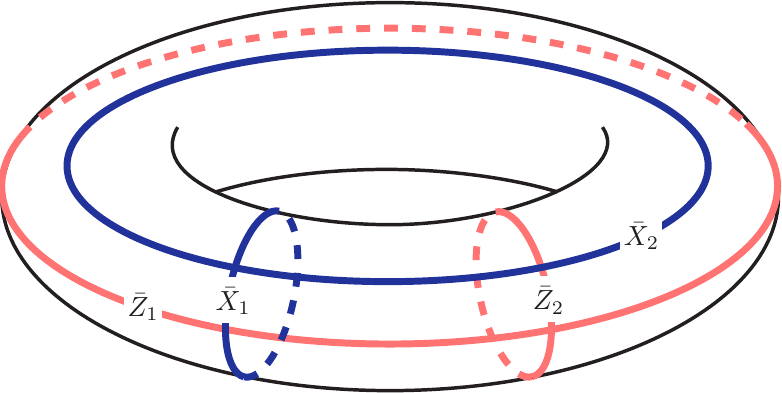}}
\caption{Logical operators that are used to define two logical qubits for a $\mathbb{Z}_2$ surface code on a torus.}\label{fig:torus}
\end{figure}

We can also apply the above procedure to the case of a $\mathbb{Z}_2$ surface code defined on a higher genus
surface. On a genus $g$ surface, the $\mathbb{Z}_2$ surface code has $4^g$ states. The logical Pauli operators
for the case of a torus are shown in Fig. \ref{fig:torus}.

Using a CAT state, we can then fault-tolerantly measure in the $\bar Y$ basis, allowing full implementation of the
Clifford group for logical qubits encoded using the genus. For example, on a torus, we have two logical qubits, one of
which can be used as a logical ancilla. Utilizing the CAT state for fault-tolerant measurements in the circuits described in Sec. \ref{sec:circuits},
we can implement all single-qubit Clifford operations. On higher genus surfaces, we can therefore implement the full Clifford group by using one of the
logical qubits as a logical ancilla.

\subsubsection{Application to 3D surface codes}

\begin{figure}
\centerline{\includegraphics[width=0.4\textwidth]{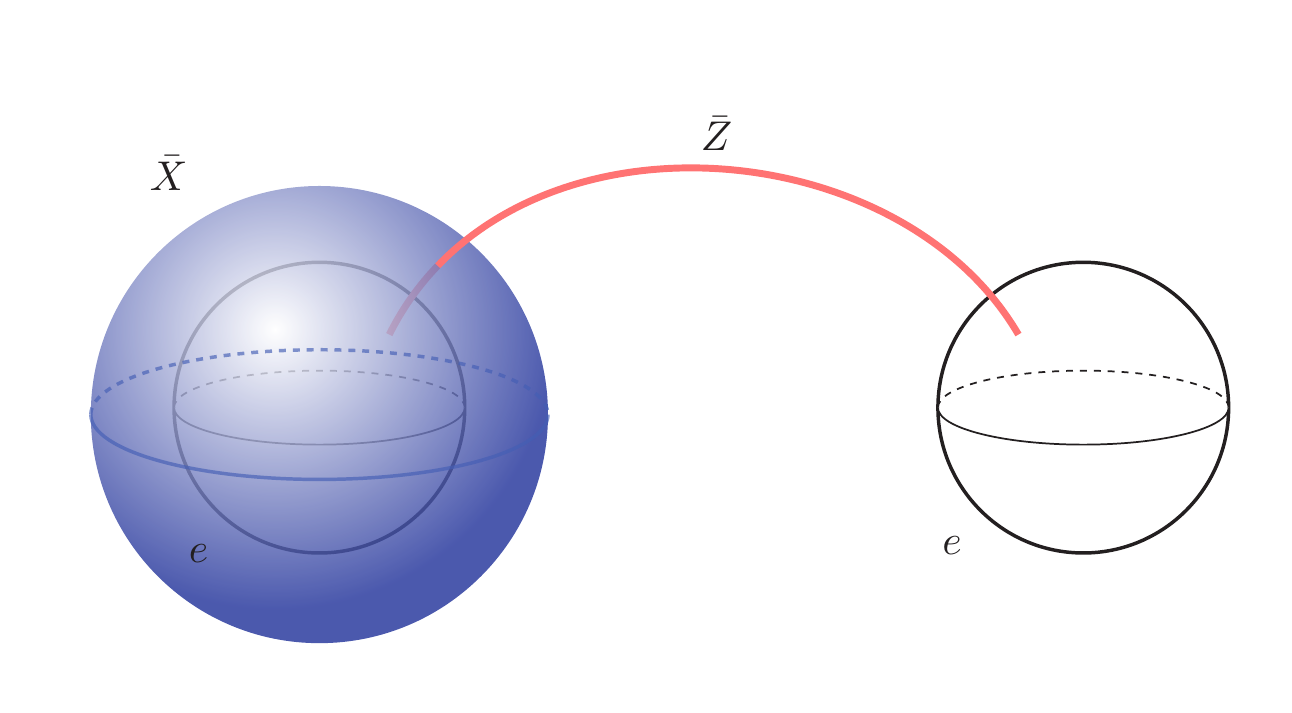}}
\caption{Encoding a logical qubit using two spherical holes in three dimensions. While the logical $Z$ operator is still given
by a Pauli-$Z$ string connecting two spheres, the logical $X$ operator is defined as a Pauli-$X$ shell covering the spherical hole. }\label{fig:hole3D}
\end{figure}

It is interesting to note that this method of performing Clifford gates can be straightforwardly extended to surface codes in higher dimensions as well.
In 3 dimensions, the analog of the hole encoding is a sphere encoding, where we consider two far-separated spheres where the
$e$ particle is condensed on the surface of the sphere. The logical $\bar X$ and $\bar Z$ operators are as shown in Fig. \ref{fig:hole3D}.
The measurement of $\bar Y$ could then be achieved with the help of a CAT state consisting of $k$ qubits placed along
the support of the $\bar Y$ operator.

If $p\, d^2 \ll 1$, where $p$ is the single qubit error probability,  the measurement of graph operators with CAT states require a time overhead $\mathcal{O}(d^2)$; the preparation of the CAT states
requires $\sim d$ steps, while the measurement of $\bar Y$ must be performed $\sim d$ times to be fault-tolerant. Furthermore, there is an
additional space overhead due to the $\sim d$ physical qubits required for the CAT states. (In 3 dimensions this space overhead
is $\mathcal{O}(d^2)$). However, if $p\, d^2$ is of order one or higher, then one needs to repeat the measurement exponentially many times in $d$
to make it fault tolerant. See Appendix \ref{catapx} for more detail. Although CAT state measurements can be useful for small codes,
it becomes impractical for large code distances. In the next section we introduce another method to perform the required joint
measurements to circumvent this problem.

\subsection{Twist defect ancilla}
\label{sec:jointsurface}

Here we explain how the circuits discussed in Sec. \ref{sec:circuits} can be implemented in surface codes
without the use of CAT states. We will see for any encoding of logical qubits, as long as we include a logical
ancilla encoded with bulk twist defects, we can carry out all of the fault-tolerant measurements required for the circuits in
Sec. \ref{sec:circuits} .

\begin{figure*}
\centerline{\includegraphics[width=0.8\textwidth]{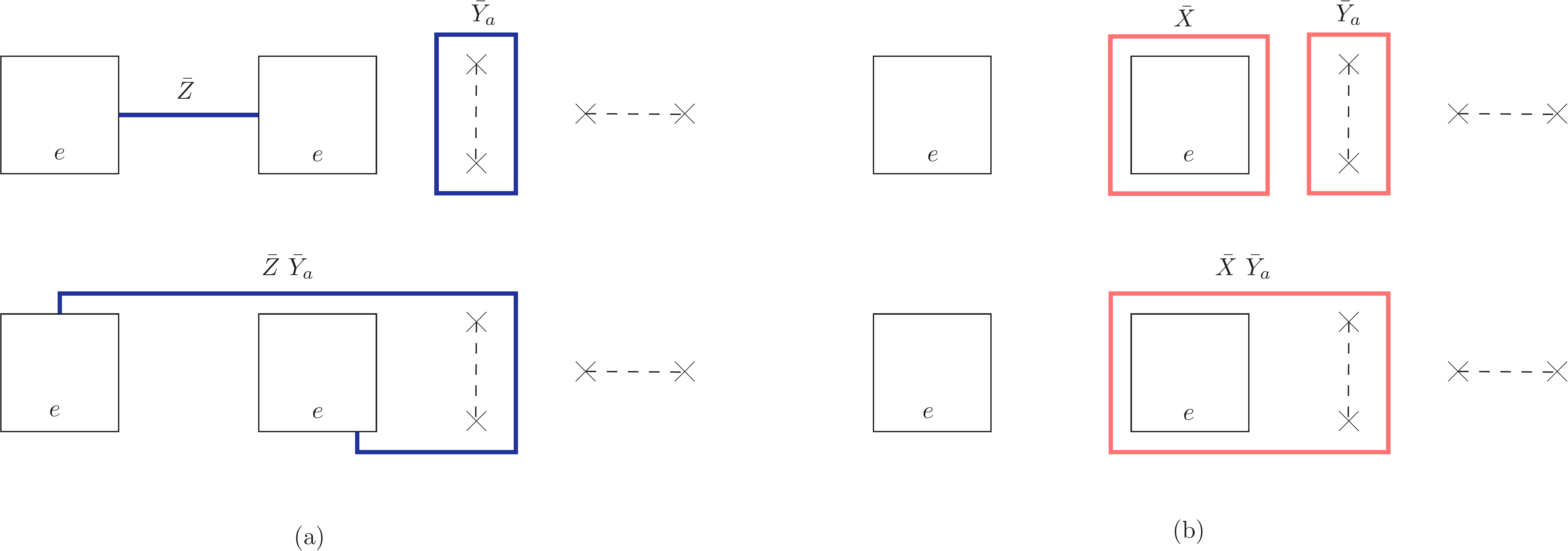}}
\caption{Finding the string corresponding to parity operator measurements in a hole based encoding. Here a pair of $Z$-cut holes are used to encode a logical qubit. There is also an ancilla qubit encoded in four twist defects. Logical operators of the ancilla qubit are specified by subscript $a$. In (a) (top panel), one can see the string operators corresponding to $\bar Z$ and $\bar Y_a$ separately. Both of them are given by a product of Pauli $Z$ operators acting on individual physical qubits. If we want to measure the parity operator $\bar Z~ \bar Y_a$, we can deform the string corresponding to $\bar Z$ in such a way to overlap the left side of the string corresponding to $\bar Y_a$. Since $\sigma_Z^2=1$ the overlapping part cancels out, and we would get the connected string shown in the lower panel of (a). In (b) we illustrate the string associated with the parity operator $\bar X \bar Y_a$. Since the logical qubit is encoded in a pair of $Z$-cut holes, the $\bar X$ operator is given by a Pauli-$X$ string operator encircling one of the two holes. To do the joint measurement, we use a Pauli-$X$ string encircling the two twists to represent $\bar Y_a$ shown in (b) up; As we explained, it is equivalent to the Pauli $Z$ string encircling the same twists which we used for the $\bar Z~ \bar Y_a$ measurement. By deforming the $\bar X$ string and making an overlap with the $\bar Y_a$ string, the shared part cancels out and we get the string shown in the lower panel of (b) for the $\bar X~\bar Y_a$ operator.}\label{fig:surfaceJM}
\end{figure*}

Here we are going to explain in the context of a simple example how a twist defect ancilla allows the required fault-tolerant joint measurements.
Suppose that we have some string operator running through some patch of a surface code,
corresponding to a logical operator $O$. It can be a non-trivial loop encircling a hole or some twist defects, or a string that connects
two same-type edges in a planar code. Also assume we have a logical qubit encoded in four twist defects in the bulk of the same patch.
Imagine we want to measure the parity operator $\bar{O}\, \bar{Y}_a$. One can get a simple string corresponding to this operator by deforming the
string corresponding to $\bar{O}$ in a way to also encircle the pair of twists that $\bar{Y}_a$ encircles. Now, if we measure this new single string
operator, using the usual procedures used to measure string operators fault tolerantly, we would get the parity value, without measuring
each individual logical operator separately. The same procedure works if one wants to measure other logical parity operators like $\bar{O}\,\bar{X}_a$
and $\bar{O}\,\bar{Z}_a$; one just needs to deform the string associated with $\bar{O}$ so it encircles the correct pair of twists. The explicit implementation
of this procedure when $\bar{O}$ is the logical $X$ or $Z$ operator of a qubit encoded in a pair of $Z$-cut holes is shown in Fig. \ref{fig:surfaceJM}.

Having the tools, implementing each protocol is quite easy. We only explain the $S$ gate implementation in the context of the hole based encoding,
but the procedure is essentially the same for other gates (CNOT and $SHS$) in other encoding schemes.

\begin{figure*}[ht]
    \centering
    \includegraphics[width=0.8\textwidth]{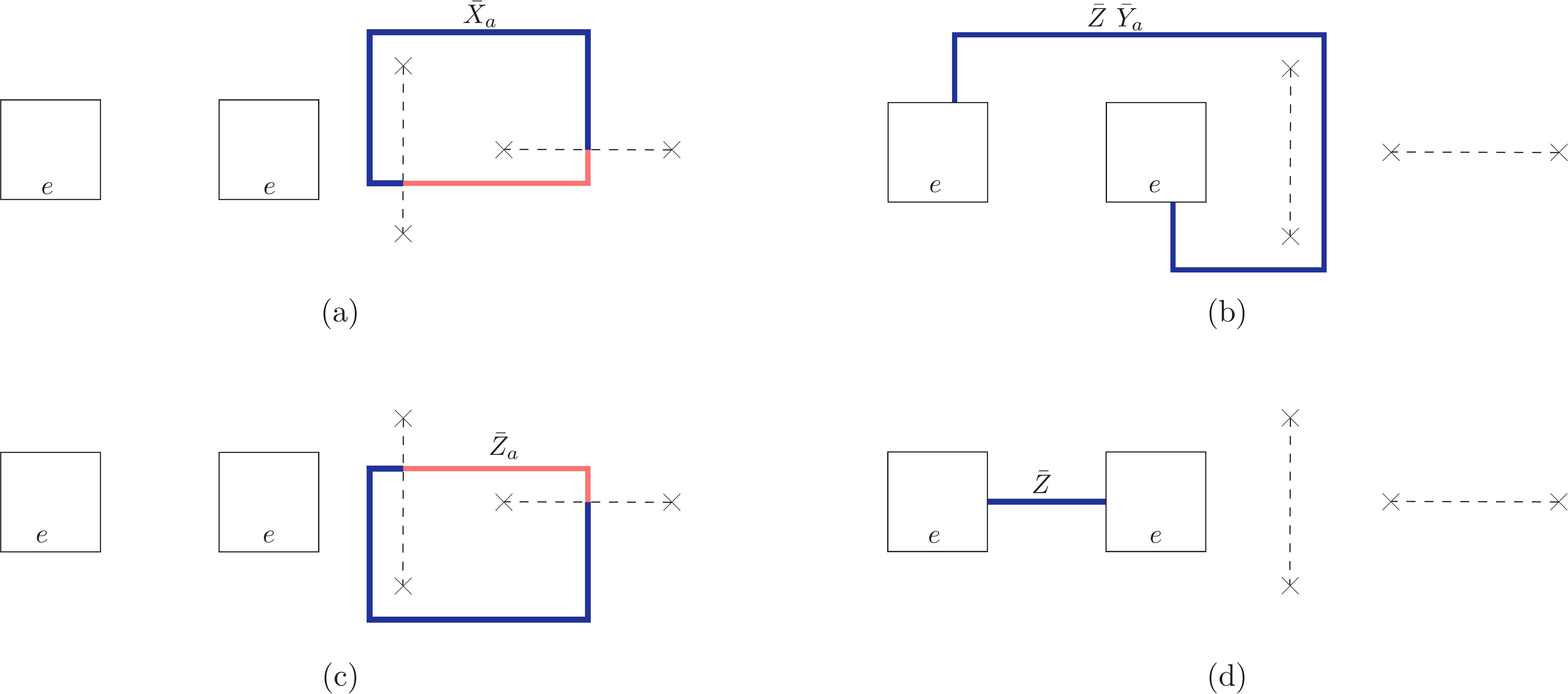}
    \caption{(Color online) $\bar S$ gate implementation in hole encoding using a twist defect as ancilla. a) Initialize the ancilla in $\ket{+}$ state, using the string operator shown above. The red (light gray) and blue (dark gray) parts correspond to $X$ and $Z$ Pauli strings respectively. b) Measure parity operator $\bar Z~ \bar Y_a$ using the Pauli-$Z$ string shown above. After measurement, glue the patch together by doing $d$ rounds of error correction with full stabilizers. c) Measure $\bar Z_a$ for ancilla qubit, using the string operator shown above and again glue the surface together. d) If the results of two measurements in part b and c had different signs, apply $\bar Z$ transversally.}\label{fig:surfaceSgate}
\end{figure*}

Assume we have a logical qubit $\ket{\psi}$ encoded in a pair of $Z$ cut holes and we want to apply the $S$ gate to it. Assume we
have also an ancilla qubit nearby encoded in four twist defects. The following is the step by step description for implementing the $S$ gate:
\begin{enumerate}
\item
Prepare the ancilla qubit in the $\ket{+}$ logical state (Fig. \ref{fig:surfaceSgate}a).

\item
Measure the $\bar{Z} \, \bar{Y}_a$ string operator shown in Fig. \ref{fig:surfaceSgate}b using string measurement method explained in
Sec. \ref{sec:stringmeasurement}. After reading the measurement result, turn on all the stabilizers and run $d$ rounds of error correction to re-attach the lattice.

\item
Measure the $\bar{Z}_a$ string shown in Fig. \ref{fig:surfaceSgate}c. Again after doing the measurement, turn on all the stabilizers and go through $d$ rounds of error correction.

\item
If the results of two measurements had the same sign, the logical qubit has been projected in the $\bar{S}\ket{\psi}$ state and the procedure has been finished.
Otherwise, perform a transversal phase flip $\bar{Z}$ gate along the string shown in Fig. \ref{fig:surfaceSgate}d to get the desired result.

\end{enumerate}

The procedures for implementing $\bar{S}\bar{H}\bar{S}$ and $\overline{\text{CNOT}}$ are quite similar to what is described above.
One needs only to choose the right string for the measurement, and re-attach the lattice together after each measurement by going through $d$ rounds of error correction.

\begin{figure*}[ht]
    \centering
    \includegraphics[width=0.7\textwidth]{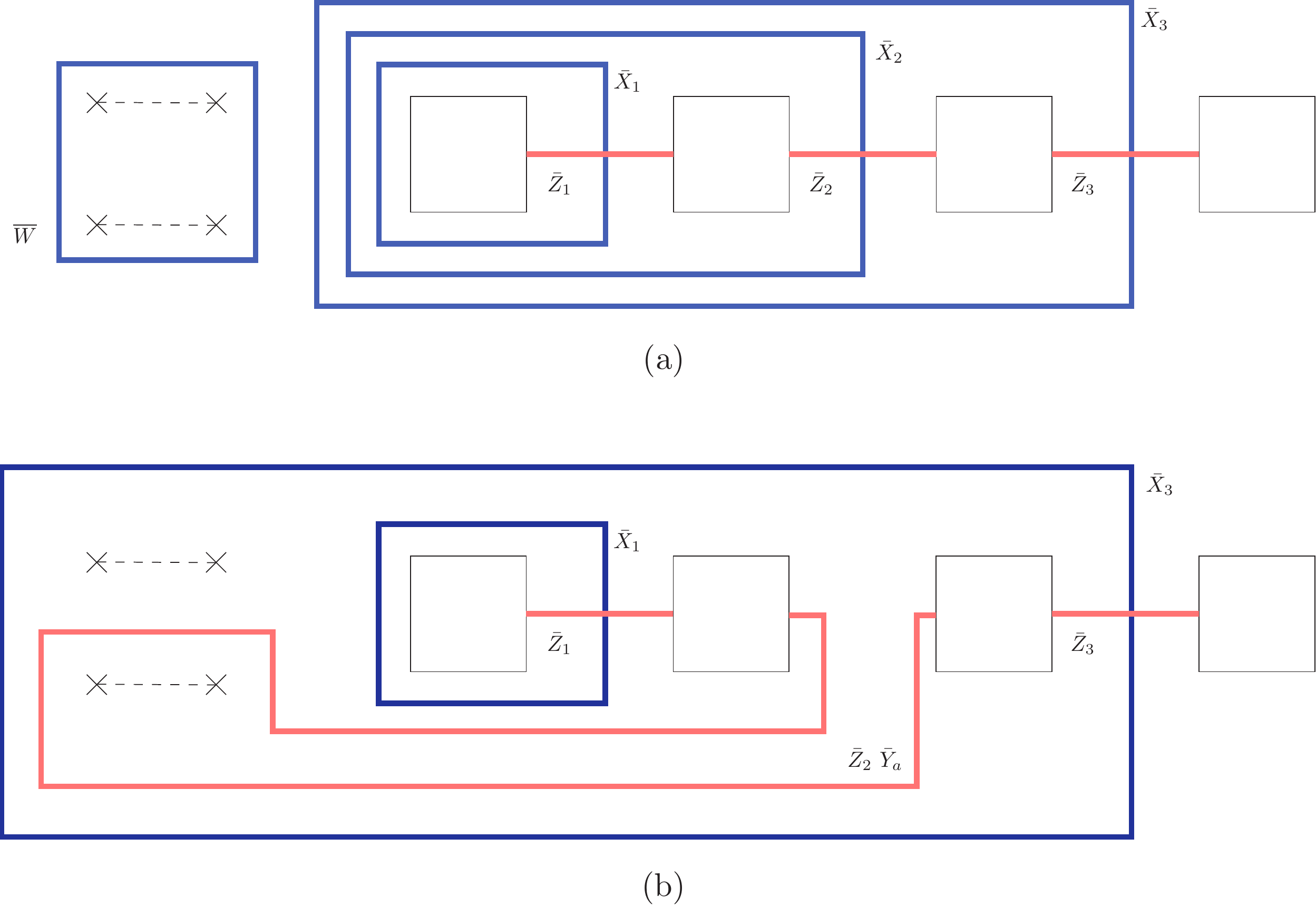}
    \caption{a) One can define $n-1$ logical qubits using $n$ holes in the dense encoding by utilizing the logical operators shown in the figure.
Four twist defects on the left are used as an ancilla qubit to implement Clifford gates. The idle string $\bar W$ contains no logical information
and will be initialized to $+1$. b) Joint measurement is performed by deforming the logical strings to encircle relevant twist defects.
Note that the $\bar W$ string is used to expand the $\bar X_3$ operator to protect its information during the $\bar Z_2 ~\bar Y_a$ string measurement.}\label{fig:denseHole}
\end{figure*}

Note that the same techniques can also be used for the dense hole encoding, where a logical qubit is defined
per each hole instead of two holes. Fig.~\ref{fig:denseHole} illustrates the way logical operators are defined
in the dense encoding as well as a typical string used for joint measurements. However the dense encodings will not
necessarily be advantageous for space overhead, for the following reason.

Since the string measurement creates new edges in the system, it can potentially reduce the code distance.
One should keep this in mind when performing joint measurements. This issue becomes more pronounced
if one needs to measure long string operators, for example in dense encodings (see Fig. ~\ref{fig:denseHole}) ,
or to perform long range CNOTs even in sparse encodings. In such cases, the string may
pass too close to many other logical qubits encoded in the patch.

To avoid this problem during the string measurements, it is always possible to encode the qubits far enough away from each other. However
this could be inefficient since usually it increases the spatial overhead considerably. Other workarounds may be possible
in certain cases that will result in no or very small increases in spatial overhead. As an example, we will explain how one
can address this issue in sparse hole and dislocation encodings.

\begin{figure*}
\centerline{\includegraphics[width=0.8\textwidth]{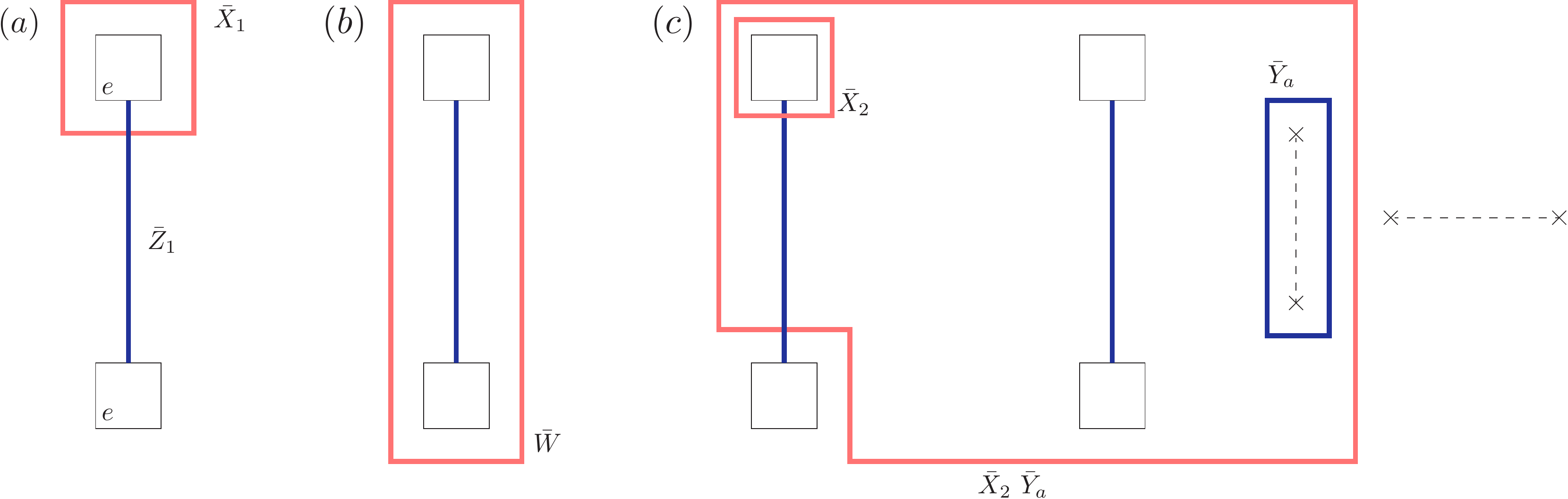}}
\caption{ A patch of holes encoding four $X$-cut logical qubits and an ancilla qubit encoded using twist defects. a) Each pair encode a logical qubit with $\bar Z$ connecting two holes and $\bar X$ encircling the top hole. b) For each pair of holes, the Pauli-$X$ string $\bar W$ that encircles both holes in topologically non-trivial but contains no information. We can initialize all such idle strings to $+1$ and use them to extend other strings that need to be measured to maintain the code distance. Note that the Pauli-$Z$ string that encircles the holes can be absorbed into hole boundaries and so is always equal to $+1$. c) How one can utilize idle strings to perform long range joint measurements.}\label{fig:holearrangement}
\end{figure*}

When a hole based encoding is used, typically one places holes on a square lattice with distance $d$ as is depicted in Fig. \ref{fig:holearrangement}, and each pair stores one qubit of information as usual (Fig.~\ref{fig:holearrangement}a). However, there are still some nontrivial loops, like the string $\bar W$ in Fig.~\ref{fig:holearrangement}b, which are not used to encode any information. We call these strings \textit{idle strings} and utilize these unused degrees of freedom to perform long range string measurements without decreasing the code distance. The idea is that we first initialize all idle strings encircling logical qubits to $+1$ and use them to extend the other strings through the code patch. For example, as is shown in Fig.~\ref{fig:holearrangement}c, to perform the joint measurement $X_2 Y_a$, where $X_2$ and $Y_a$ are plotted in the figure, one can use the string that also encircles the logical qubit in between, without affecting the measurement result, which in turn helps to keep the code distance $d$. More details can be found in the caption.

\begin{figure*}[t]
\centerline{\includegraphics[width=\textwidth]{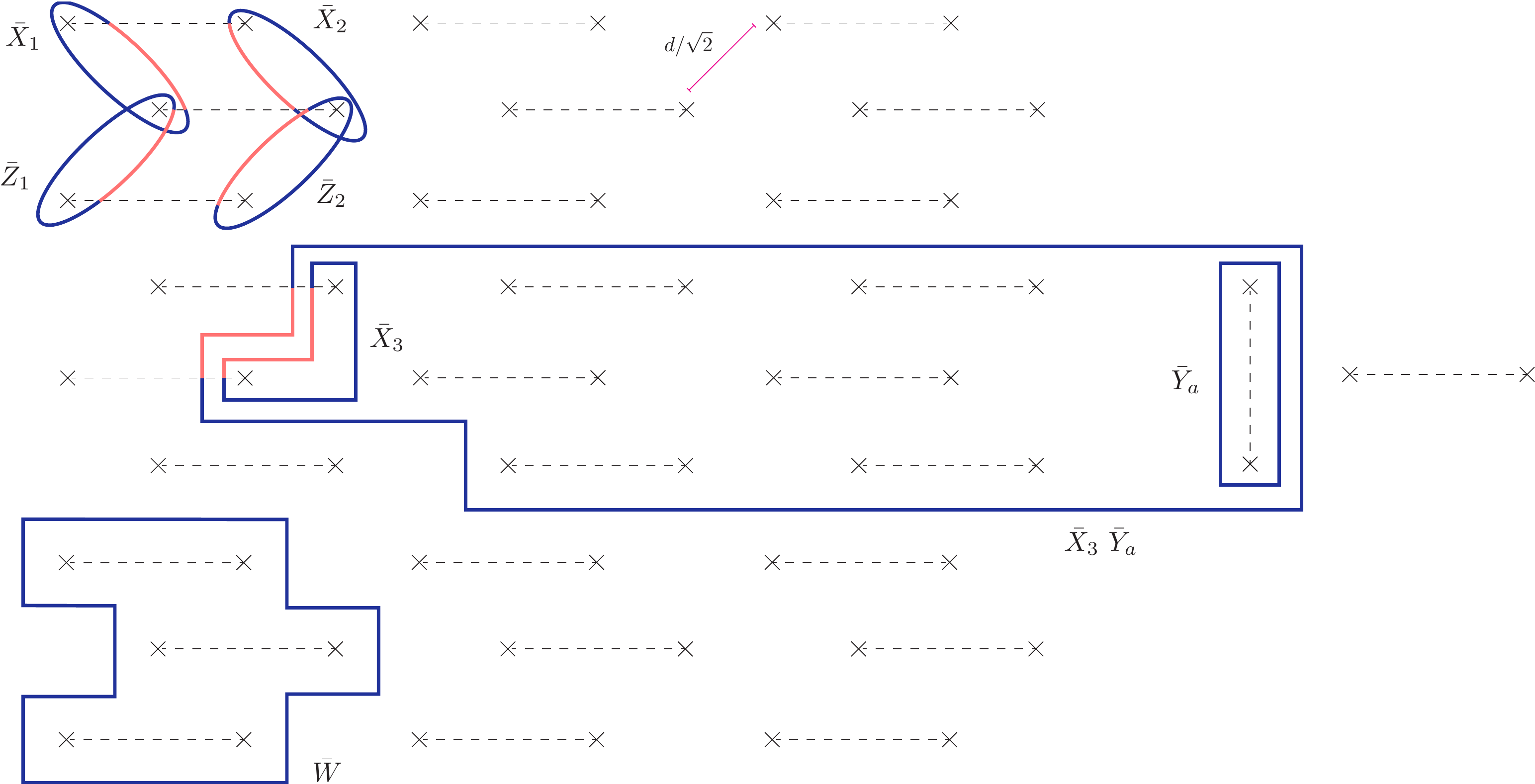}}
\caption{(Color online) Typical arrangement of twist defects on an underlying rotated square lattice with lattice size $d/\sqrt{2}$ which results in a dislocation code with distance $d$. We use $3$ twist defects to encode one logical qubit. The strings that encircles two logical qubits that are encoded in adjacent twist defects, like $\bar W$, are non-trivial but contain no information. We initialize all these idle strings to $+1$ to use them for extending measurement strings. Note that the Pauli-$X$ string (red (light gray)) that traces the same path as $\bar W$ is equivalent to $\bar W$. It is enough to initialize just one of them. c) A long range $\bar X_3 \bar Y_a$ measurement that maintains the code distance using idle strings. The string passes with at least $d/4$ distance from twist defects. This alongside using idle strings ensures during measurement no short error string could occur.  }\label{fig:dislocationarrangement}
\end{figure*}

A similar idea can be used in dislocation codes. To have a dislocation code with distance $d$ one can arrange twist defects on a \textit{rotated} square lattice with lattice size $d/\sqrt{2}$ as illustrated in Fig.~\ref{fig:dislocationarrangement}. We can use three twist defects to encode a single logical qubit. In this way for each two logical qubits, there would be a non-trivial idle string that encircles the six twist defects and contains no information. We can initialize these strings to $+1$ and use them to perform long range string measurements without reducing the distance in a similar way to the hole encoding case. As an example, a long range $\bar X_3~\bar Y_a$ measurement between a logical qubit and an ancilla qubit is shown in Fig.~\ref{fig:dislocationarrangement}. Note that after measuring the shown string the code  will divide into two patches, each protected by distance $d$.

For the reasons discussed above, dense encodings do not appear to be more advantageous than sparser encodings,
in the limit of a large number of logical qubits arranged in a two-dimensional space. On the other hand, for small
numbers of logical qubits or logical qubits placed along a line, the dense encodings do have improved spatial overhead
than sparser encodings.

\subsubsection{Classical tracking of single qubit gates}

As was mentioned in Sec.~\ref{sec:classicalTracking}, instead of applying single qubit gates in a quantum circuit,
one can trade CNOT gates in that circuit for conjugated versions of them and modify the final measurements.
But this will be useful only if one can implement conjugated versions of CNOT with almost the same number of
steps as the CNOT itself. Let us consider the $\bar{S}^\dagger~\overline{\text{CNOT}}~\bar{S}$ circuit (Fig.~\ref{fig:SCNOTS}) as an example.
The only non-trivial part of that circuit is the $M_{\bar{X}_a \bar{Y}_t}$ measurement, since this time the $Y$ operator appearing
in the operator to be measured is associated with a logical data qubit, whose encoding is arbitrary.
For the dislocation code this is clearly not an issue\cite{hastings2014} because we can measure the logical qubits in
any Pauli basis fault tolerantly and hence the same joint measurement techniques described here can be utilized to
measure the $\bar X_a \bar Y_t$ parity operator.

\begin{figure}
\centerline{\includegraphics[width=0.5\textwidth]{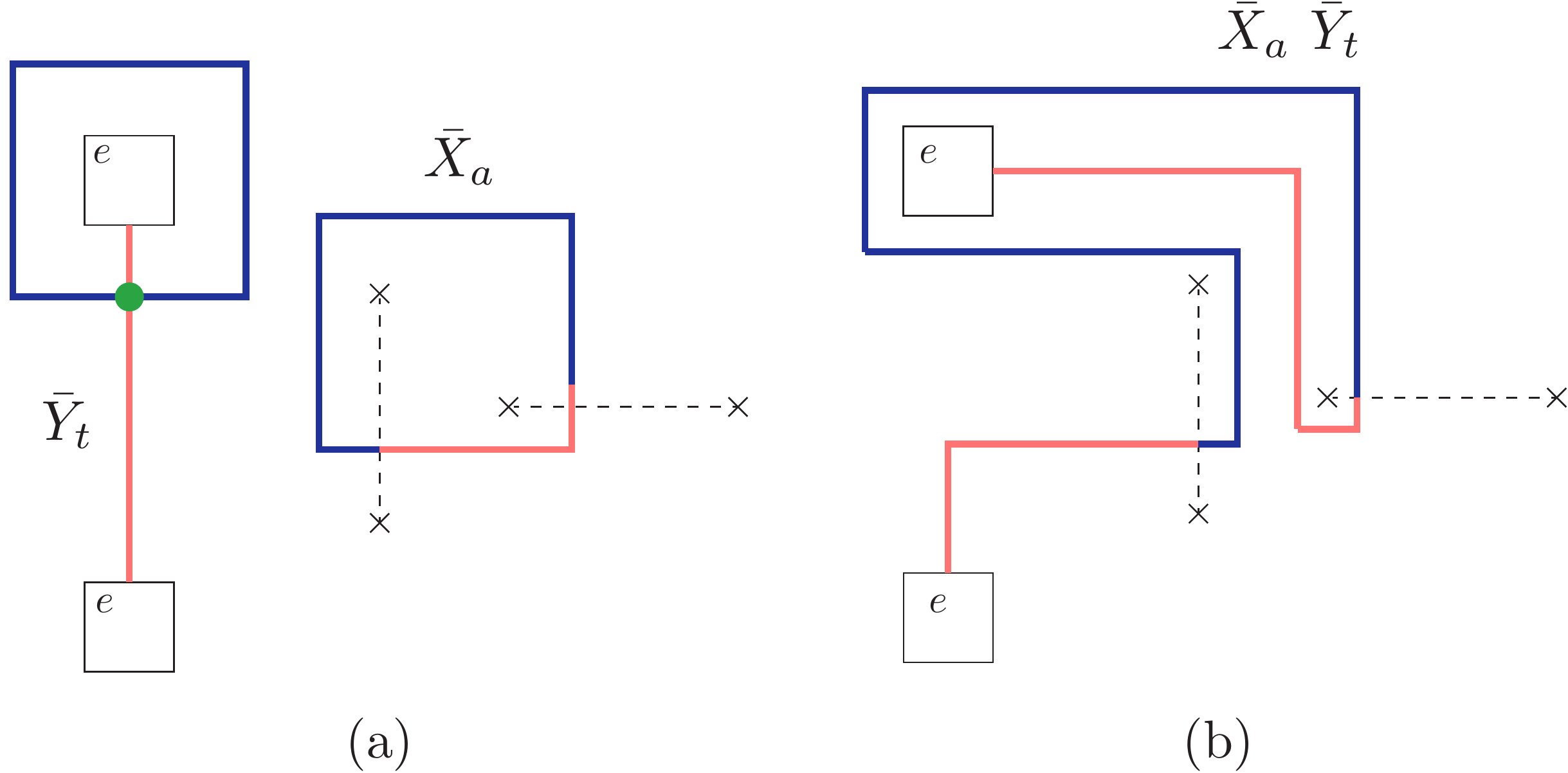}}
\caption{a)  $\bar X_a$ and $\bar Y_t$ operators. b) $\bar X_a \bar Y_t$ operator given by a simple string.}\label{fig:holeEncoding_CK}
\end{figure}

If the logical qubits are encoded using other types of defects, one needs to find a simple string (as opposed to graph) representation
for the parity operator. Remarkably, this can be done in any encoding scheme, as long as there exist ancilla qubits in the twist defect encoding.
An example is shown for the case of hole encoding in Fig.~\ref{fig:holeEncoding_CK} where one can find a simple string representation for the parity operator
$\bar{X}_a \bar{Y}_t$. To identify the logical operators in Fig.~\ref{fig:holeEncoding_CK} (a) and (b), we have used the fact that the double
loop around a single twist defect is a logical identity (see Fig.~\ref{fig:trivialDoubleLoop} ). The $Y$ measurement at the end of the modified
quantum circuits can also be done similarly, by initializing the ancilla qubit in $\ket{+}$ state and then measuring the $\bar X_a \bar Y_t$ operator.

.

\section{Hyperbolic code}
\label{sec:hyper}

The hyperbolic code is another variant of the surface code that uses different tilings of 2D surfaces to improve the
encoding rate\cite{breuckmann2016}. Although hyperbolic codes are very efficient for storing information, no
procedure is known for fault-tolerant implementation of the Clifford group. Ref. \onlinecite{breuckmann2017}
proposes two possibilities for quantum information processing: (1) to perform Dehn twists, which can be
used to either move the qubits around in storage, or to perform a logical CNOT between qubits stored in the same handle,
and (2) to use lattice surgery to convert encoded information to a surface code, perform the necessary computations, and convert back
to the hyperbolic code.

In this section we demonstrate how our methods can be used to implement fault-tolerantly the full Clifford gate set directly within the hyperbolic code,
without moving the information into another quantum code patch.

\begin{figure}[h]
\centerline{\includegraphics[width=0.5\textwidth]{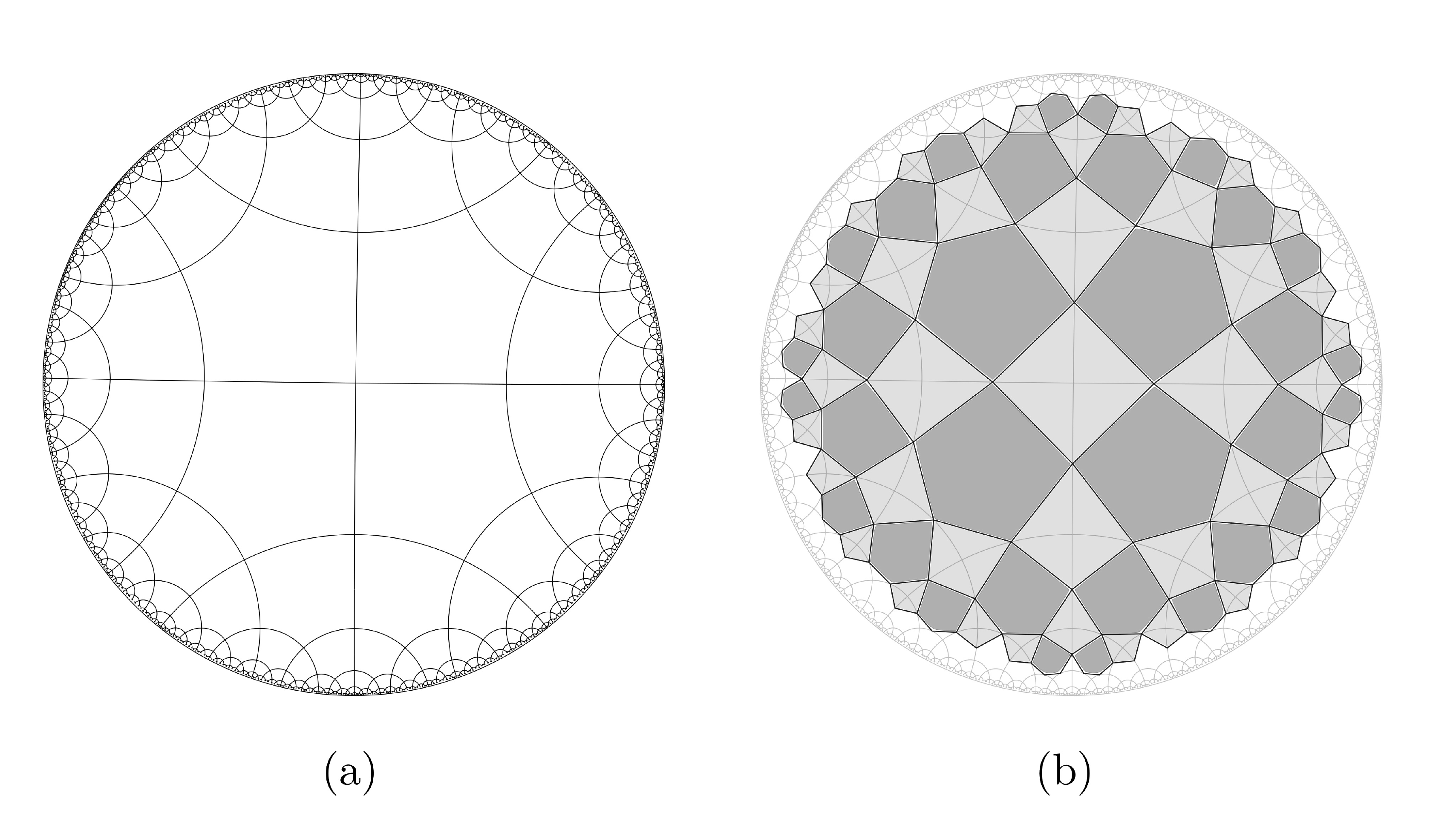}}
\caption{a) $\{5,4\}$ tiling of the hyperbolic plane. b) The rectified tiling r$\{5,4\}$ can be constructed by connecting the midpoints of the edges in
the $\{5,4\}$ tiling. The original $\{5,4\}$ tiling is also shown with light solid lines for comparison }\label{fig:hyper}
\end{figure}

The hyperbolic code is based on a tiling of a closed surface with regular polygons. A specific tiling is described by a set of two numbers $\{p,q\}$,
known as Schl\"{a}fli symbols, which represents a tiling of the plane with regular $p$-sided polygons such that $q$ of them meet at every vertex.
On a Euclidean plane, internal angles of a regular $p$-sided polygon are equal to $(p-2)\pi/p$.
On the other hand if $q$ polygons are to meet at a vertex, the internal angles should be equal to $2\pi/q$. Comparing these two, one can
see that only tilings with $1/p+1/q=1/2$ can be realized on the Euclidean plane. However, one can use hyperbolic surfaces --
surfaces with constant negative curvature --  to realize $\{p,q\}$ tilings  with $1/p+1/q<1/2$, since the sum of the internal angles
of a regular polygon on a hyperbolic plane is less than $(p-2)\pi$. Fig. \ref{fig:hyper}a illustrates the $\{5,4\}$ tiling of the hyperbolic plane.

Given a $\{p,q\}$ tiling, one can define a stabilizer code where physical qubits lie on the edges and each vertex (plaquette) represents a
$Z$-type($X$-type) stabilizer. A topologically non-trivial closed hyperbolic surface with $g$ handles has $2g$ non-trivial independent
loops which can be used to define $2g$ logical qubits that are stabilized by the code. For large distances
and fixed number of physical qubits, hyperbolic codes can encode more logical qubits compared to normal surface codes. However in order
to realize such codes in an experimental system that is constrained to the Euclidean plane, non-local interactions are required.

If one prefers to work with a form similar to the surface code which was described in Section \ref{sec:rev}, where qubits lie on the lattice sites and
all stabilizers are given by plaquette operators, one can use the \textit{rectified} lattice, denoted by r$\{p,q\}$, which is constructed by
connecting the midpoints of the edges in a $\{p,q\}$ lattice(Fig. \ref{fig:hyper}b). The rectified lattice tiles the plane with regular
$p$-sided and $q$-sided polygons. In this new form, qubits lie on the vertices and $p$ and $q$ sided plaquettes represent $X$ and $Z$ stabilizers respectively.

\begin{figure}[h]
\centerline{\includegraphics[width=0.5\textwidth]{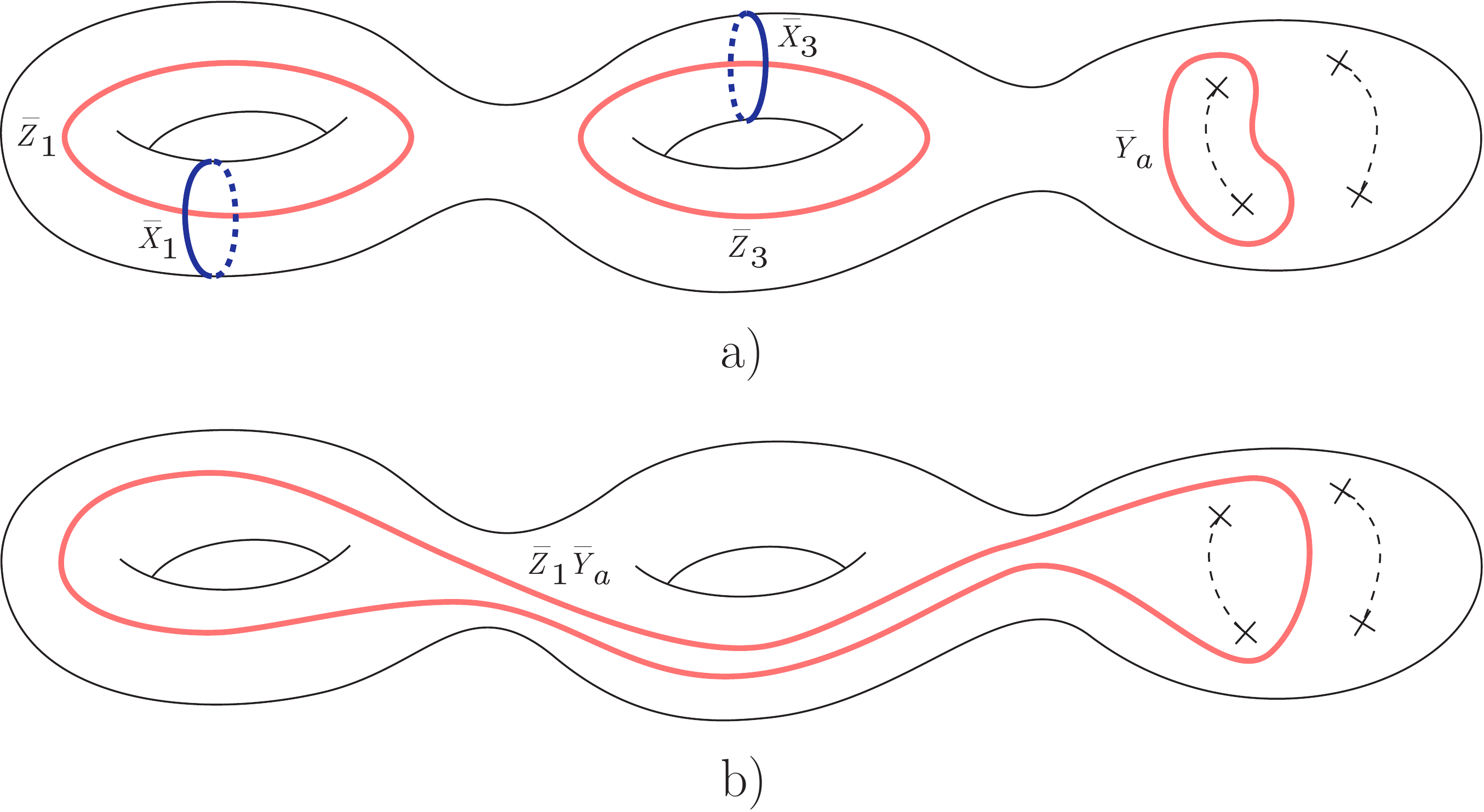}}
\caption{a) Hyperbolic surface with $2$ handles (genus $g =2 $) encoding $4$ logical qubits and four twist defects
to encode the ancilla qubit. Logical qubits are encoded by using non-trivial loops on the surface. Some logical operators
are shown in the figure as examples. The $\bar X_2$ and $\bar X_4$ ($\bar Z_2$ and $\bar Z_4$) operators are
given by Pauli-$X$(Pauli-$Z$) strings that trace out the same loops as $\bar Z_1$ and $\bar Z_3$ ($\bar X_1$ and $\bar X_3$)
operators respectively. b) A sample two qubit parity operator. }\label{fig:HyperDefect}
\end{figure}

The joint measurement circuits for implementing Clifford gates can be used in the hyperbolic codes as well.
As in the discussion of the surface code in Sec. \ref{sec:surface}, we can again consider a set of ancilla qubits that comprise a CAT state in order
to help us measure operators that consist of $\bar Y_a$.

Alternatively, as in the surface code discussion, we do not need any CAT states if we
use bulk twist defects to encode a logical ancilla. One can use the
original $\{p,q\}$ lattice and create defects in the bulk by following a procedure similar to what was described in Section \ref{sec:dislocationencoding}.
However one should be careful not to decrease the code distance and to keep track of what happens to other logical qubits.
A more straightforward approach would be to select an arbitrary plaquette, divide it into a $2d\times 2d$ square lattice and create
 a pair of dislocation lines to encode the logical ancilla qubit. Dividing a plaquette by a square lattice clearly does not change the topology
of the surface and keeps the code distance fixed.

Having a logical ancilla qubit encoded with twist defects in hand, performing joint measurements and implementing the
quantum circuits described in Section \ref{sec:circuits} is straightforward. The single and two qubit parity operators used
for Clifford group gates would be given by simple Pauli strings running through the hyperbolic plane (Fig. \ref{fig:HyperDefect}a) .
Using the string measurement method, these operators can be measured fault-tolerantly using $d$ rounds of error correction.
Fig. \ref{fig:HyperDefect}b shows a typical Pauli string representing a two-qubit parity operator.

There is another variant of the hyperbolic code, the semi-hyperbolic code \cite{breuckmann2017}, where one divides all polygons of
a $\{p,q\}$ tilling by a $l \times l$ square lattice. For large $l$ the code would be essentially a normal surface code placed
over a topologically non-trivial surface and all efficiency of the hyperbolic construction would be lost. However, it would
improve the error threshold of the code \cite{breuckmann2017}. The optimal $l$ should be chosen according to this trade-off.
Our joint measurement scheme can be straightforwardly applied to the semi-hyperbolic codes as well, in the same way as the hyperbolic codes.

\section{Color code}
\label{sec:color}

Color codes are another form of 2D topological codes, with the advantage of higher encoding rates and also allowing for
natural transversal logical operations on the qubits. However, color codes usually have smaller error thresholds
compared to surface codes. Nevertheless the trade-off between overhead and error thresholds could potentially favor the
color codes in future experiments. Although there are already known methods for fault-tolerant quantum computing with
color codes \cite{bombin2006,fowler2011,landahl2014}, here we point out that implementing the logical Clifford gates using
the joint measurement techniques of this paper could have its own advantages. Specifically, performing long-range two-qubit
gates are more efficient with our method as compared with the transversal or lattice surgery methods. In contrast to the case of the
surface code (without CAT states), our joint measurement protocols can be implemented in the color code in the case where the
logical ancilla arises from a hole-based encoding.

The color code can be defined on any three-colorable, three-valent lattice. Qubits lie on the lattice sites and each
plaquette corresponds to both $X$ and $Z$ stabilizer operators. The lattice structure ensures that all stabilizers commute with each other.
Each stabilizer violation corresponds to a particle. To label the particles we use the color and type of the stabilizer it violates. So, $r_x$ denotes a particle detected
by a $X$ stabilizer corresponding to a red plaquette.
Since we have three different color plaquettes (say red, blue and green), and each plaquette
corresponds to two stabilizers, naively there seems to be $6$ independent particles in the theory. However, it can be shown that one can
annihilate three particles of the same type and different colors with each other\cite{bombin2006}. Thus, only $4$ out of $6$ are really
independent particles. By considering composites of these $4$ types, we find that there are $16$ topologically distinct particles.
Indeed, it can be shown that the color code is equivalent (after a finite-depth local unitary transformation) to two copies of the
surface code,\cite{kubica2015} which has a total of $16$ topologically distinct particles.

Just like the surface code, topologically distinct particles always appear in pairs and each pair is connected via Pauli string operators. Since particles carry color, the string operators also would be
red, blue or green. For example, a red Pauli-$Z$ string connects two $r_x$ particles. Note that  a Pauli-$Z$ string violates
$X$ stabilizers at its ends but commutes with $Z$ stabilizers.

Similar to the two topologically distinct $e$ and $m$ boundaries in the surface code, the color code can have
$6$ topologically distinct types of boundaries, given that it is equivalent to two copies of the surface code (see Ref. \onlinecite{levin2013,barkeshli2013defect,barkeshli2013defect2}
for a classification of topologically distinct boundary conditions in topological phases). In paticular, the
color code can have red, blue and green boundaries where red, blue and green particles can condense.

As in the case of the surface code, logical qubits can be defined through boundary defects, holes, bulk twist defects, or by
having non-trivial genus \cite{bridgeman2017,fowler2011,bombin2011clifford,teo2013}.

Holes are created by simply not measuring stabilizers within some region. To create a hole with a red boundary, for example,
we consider a closed loop of red string, and stop measuring the stabilizers inside the loop. We also modify the stabilizers on the edge accordingly.
Then we have created a hole with a red boundary where red strings can start or end on it without violating stabilizers.
\begin{figure}[h!]
\centerline{\includegraphics[width=0.3\textwidth]{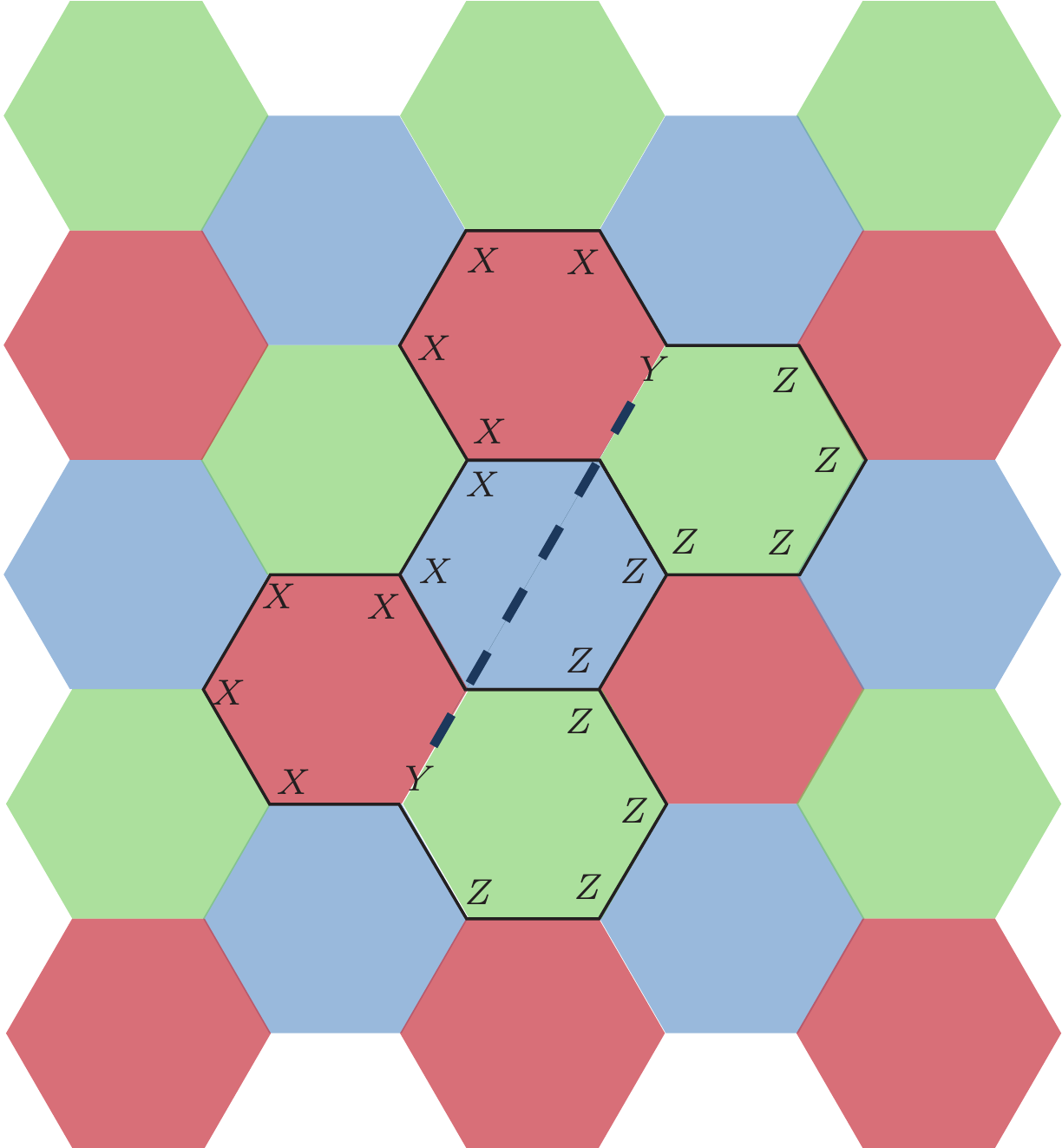}}
\caption{(Color online) Making a twist defect in the color code. To make a twist defect, we first draw the dislocation line (dashed blue line)
and remove the qubits which lie on the line. Then we merge pairs of $X$ and $Z$ stabilizers on the sides of the dislocation
line into one. Note that in color code each plaquette represents two stabilizers. In this figure half of modified stabilizers
are shown. For each stabilizer shown there is another one with $X$ and $Z$ operators exchanged. If the dislocation line
passes through a plaquette (like the blue plaquette in the middle), one should merge the $X$ and $Z$ stabilizers corresponding to the same plaquette. }\label{fig:colortwist}
\end{figure}

The procedure to create bulk twist defects in the color code is similar to the case of the surface code. One chooses a dislocation line,
removes the physical qubits over that line and merges pairs of $X$ and $Z$ stabilizers on either sides into one. Twist defects in
surface codes transform $e$ particles to $m$ particles and vice versa. Since color codes have more particles, there are more
types of twist defects one can create\cite{barkeshli2014SDG}. In Fig. \ref{fig:colortwist} we have shown one possible example of a pair of twist defects connected
to each other with a dislocation line.  The one shown in Fig. \ref{fig:colortwist} changes $r_x$ to $g_z$, $r_z$ to $g_x$, $b_x$ to $b_z$ and vice versa,
as the particles enircle the twist defect.
\begin{figure}[h!]
\centerline{\includegraphics[width=0.4\textwidth]{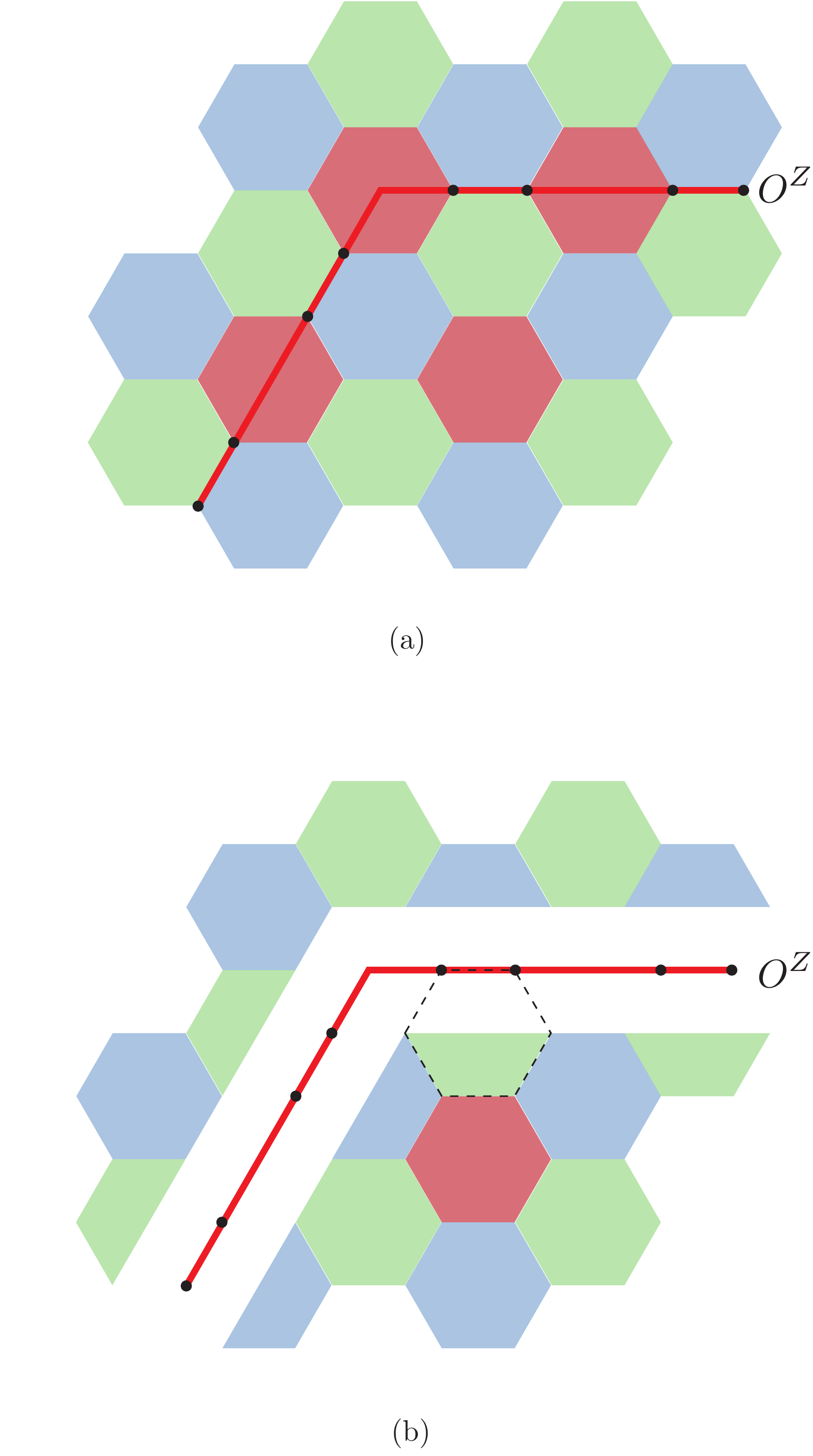}}
\caption{(Color online) String measurement in color codes. a) If we want to measure the red (dark gray) Pauli-$Z$ string operator $O^Z$, first we detach the qubits of $O^Z$ from neighboring stabilizers and turn off all red (dark gray) stabilizers in its way. b) Then we measure individual qubits in $Z$ basis and multiply the results to find the value of $O^Z$. One can correct errors by comparing the value of individual measurements and modified stabilizers with the original value of complete stabilizers. For example, the syndrome of the green stabilizer corresponding to the dashed plaquette before cutting the code, should be equal to the product of measurement outcomes of those two individual qubits and the $4$ qubit stabilizer bellow them.}\label{fig:csmeasurement}
\end{figure}

Based on the underlying encoding scheme, different methods for initialization, measurement and realization of quantum gates can be used.
Most of the techniques used in surface codes like lattice surgery and hole braiding have counterparts in color codes\cite{landahl2014,fowler2011}.
To measure a string operator, one can practically follow the same procedure used in surface codes. An example is shown in Fig. \ref{fig:csmeasurement}.
Let us say we want to measure the red Pauli-$Z$ string operator $O^Z$ shown in Fig. \ref{fig:csmeasurement}a. First, we detach from the stabilizers
the qubits that lie on $O^Z$. This will change some of the green and blue stabilizers from $6$ qubit to $4$ qubit stabilizers. We also need to turn
off all red stabilizers which $O^Z$ passes through. After the modification, the color code will look like Fig. \ref{fig:csmeasurement}b. Note that
this has effectively created a red boundary along the string, where red error strings can start and end without detection. But, just as in the case of the
surface code, these undetected errors will not change the value of $O^Z$. In the next step, we measure each individual qubit on
$O^Z$ in the $Z$ basis and also measure all stabilizers. By combining the outcome of individual measurements and modified stabilizers and
comparing them with the value of the complete stabilizers, we can detect any error that happens during the measurement process. After
correcting the errors, multiplying the outcome of individual measurements would give the value for the measurement outcome of $O^Z$.

The joint measurement circuits discussed in this paper for implementing the Clifford group can also be implemented in color codes.
If qubits are encoded in a single patch, for example using holes or dislocations, the quantum circuits
described in section \ref{sec:circuits} can be implemented using a single logical ancilla encoded with twist defects.
The rest of the protocol is directly analogous to the case of the surface code.

However, unlike the surface code, in color codes we are not restricted to use twist defects as the logical ancilla to implement
the joint measurement method. An interesting feature of color codes is that one can measure not only $X$-type and $Z$-type strings
fault tolerantly, but also $Y$-type strings. The reason is that in the color code, in contrast to the surface code, for a given plaquette
we measure both $X$ and $Z$ stabilizers and the product of these outcomes gives the value of the corresponding $Y$ stabilizer
(we need to multiply it by $(i)^n$ where $n$ is the number edges in the plaquette). This feature is a result of the fact that the color
code is a CSS code \cite{nielsen2010} constructed from two \textit{copies} of a single classical code. It is the same property
that makes transversal methods natural in this architecture. This in turn enables us to create logical qubits where
$\bar Y$ is given by a simple Pauli string, without using twist defects.

\begin{figure}[h]
\centerline{\includegraphics[width=0.5\textwidth]{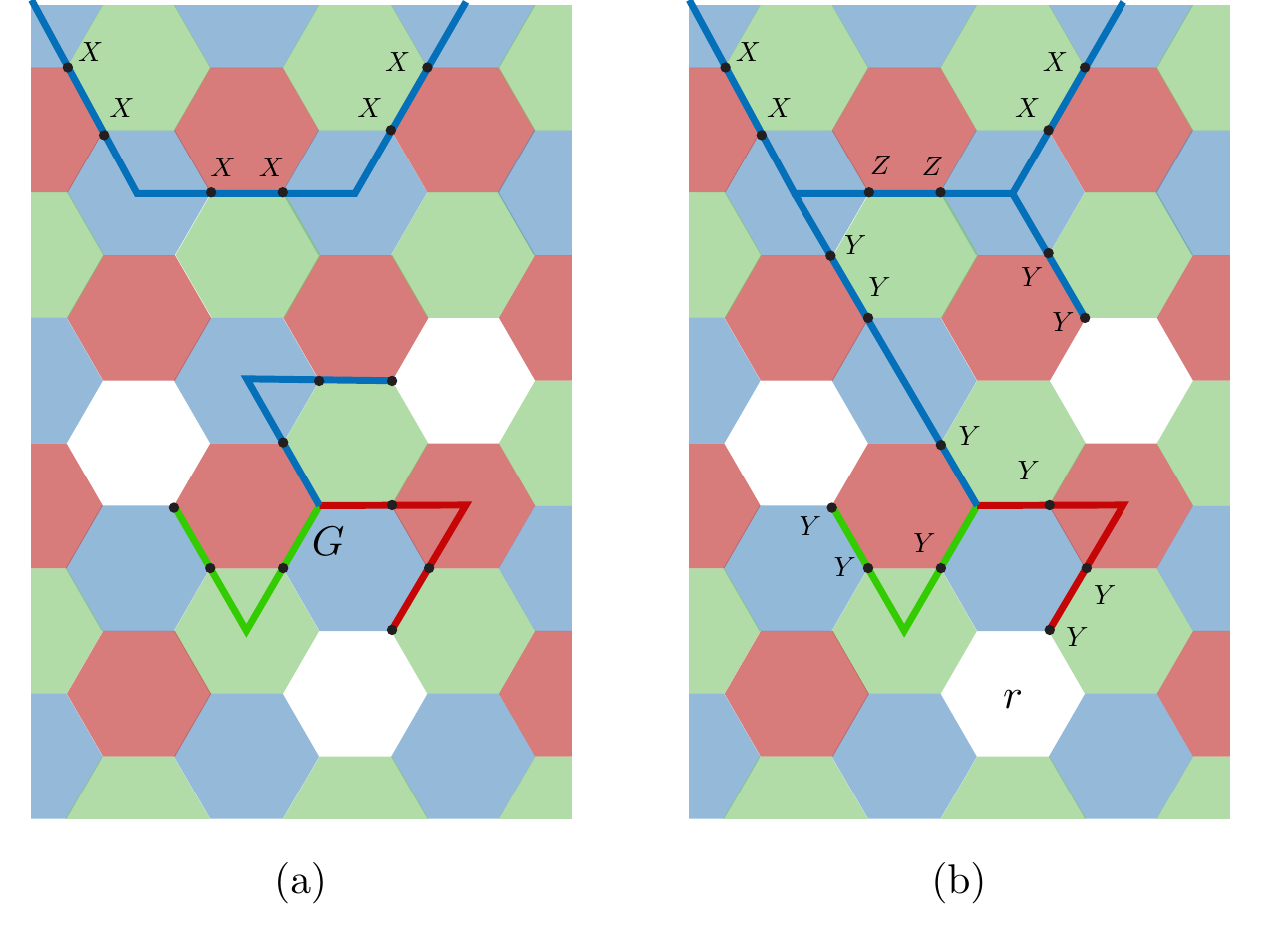}}
\caption{a) Three holes and the $G$ graph that connects them encode the logical ancilla qubit. Blank plaquettes represent holes where
we do not measure any stabilizer. $\bar X_a$, $\bar Z_a$ and $\bar Y_a$ are given by $G^X$, $G^Z$ and $G^Y$ respectively.
At the top, part of a Pauli string is shown which is a portion of $\bar X$ operator related to another logical qubit.
b) To measure two qubit parity operator $\bar Y_a \bar X$ one should deform $G$ in such a way to overlap with the
string related to $\bar X$ and measure the resulting string.}\label{fig:colorhole}
\end{figure}

To implement measurements involving $\bar Y$ with a hole encoding, we encode the logical ancilla qubit using three holes associated with
different colors, similar to the proposed hole-based encoding in Ref. \onlinecite{fowler2011}, but with a small modification.
Consider the three holes and the graph $G$ that connects them, shown in Fig. \ref{fig:colorhole}a.  We define $\bar X$  as the
$G^X$ operator, which means the product of Pauli-$X$ operators along the graph, and, similarly, the $\bar Z$  as the $G^Z$ operator.
Since the graph $G$ consists of an odd number of qubits, $G^X$ anti-commutes with $G^Z$. The advantage of this
scheme is that the logical $Y$ operator would be a Pauli-$Y$ graph operator, denoted $G^Y$ (in contrast to the proposed method in Ref. \onlinecite{fowler2011}) and can be measured fault tolerantly.

If we encode the ancilla qubit in the aforementioned three hole structure, no matter how the data qubits are encoded, as long
as the logical $X$ and $Z$ operators of the data qubits are given by deformable strings, we can use joint measurement for
quantum computation. The idea is similar to what was described in surface codes. For parity measurements, we deform
the strings to overlap and measure the resulting string (Fig. \ref{fig:colorhole}b). Since the $G$ graph has all three different colors,
we can deform it to overlap with any other string operator along a line. Then, the string measurement method can be used to
find the parity value fault tolerantly.

\section{Resource analysis}
\label{sec:resource}

In this section we discuss how our proposed methods inform the design of efficient surface code encoding schemes,
and we analyze the associated resource costs and compare them to other proposed methods of quantum computation
with the surface code.

In particular, we analyze the resource costs associated with both small and large numbers of logical qubits.
For small logical qubits, we propose new surface code schemes for encoding one and two logical data qubits and one logical ancilla qubit,
which allow implementation of the full Clifford group using our joint measurement circuits. We compare the resource
costs in terms of number of physical qubits required for a given code distance with other proposals presented in the literature.
For two logical qubits and being able to perform all Clifford group operations, our proposal is optimal; as such, we expect
it could play a useful role in near-term experimental demonstrations of fault-tolerant quantum eror correction.

For large numbers of qubits, we argue that the most optimal method to date for the surface code uses a lattice of dislocations,
and we compare the resource costs in terms of number of physical qubits for a given code distance for such an encoding
scheme with other proposed methods.

We remark that the hybrid scheme of Ref. \onlinecite{brown2017} is not included in this comparison. Although the
proposed scheme does not use state distillation for implementing Clifford gates and hence is a promising candidate
for near term surface code realization, an exact analysis of resource costs for few qubits has not been performed.
Ref. \onlinecite{brown2017} analyzed its asymptotic space overhead scaling for large $d$, although our considerations
(not summarized here) of their approach yield different results, so we leave a definitive analysis for future work.

\subsection{Single qubit codes}\label{sec:singlequbit}

Let us first consider small codes that admit implementation of all single qubit gates in the Clifford group.
In the case of color codes, we only need one logical qubit and we can implement all single qubit Clifford
gates transversally. However in some variants of surface codes, we need at least an extra logical ancilla qubit to apply
single qubit Clifford gates on logical qubits. The hole-based encoding and lattice surgery methods proposed so far
rely on state injection and distillation for implementing the full Clifford group gate set, whereas
triangle codes and the surface code with joint measurement, presented in this paper, do not.

\begin{figure}[h]
\centerline{\includegraphics[width=0.5\textwidth]{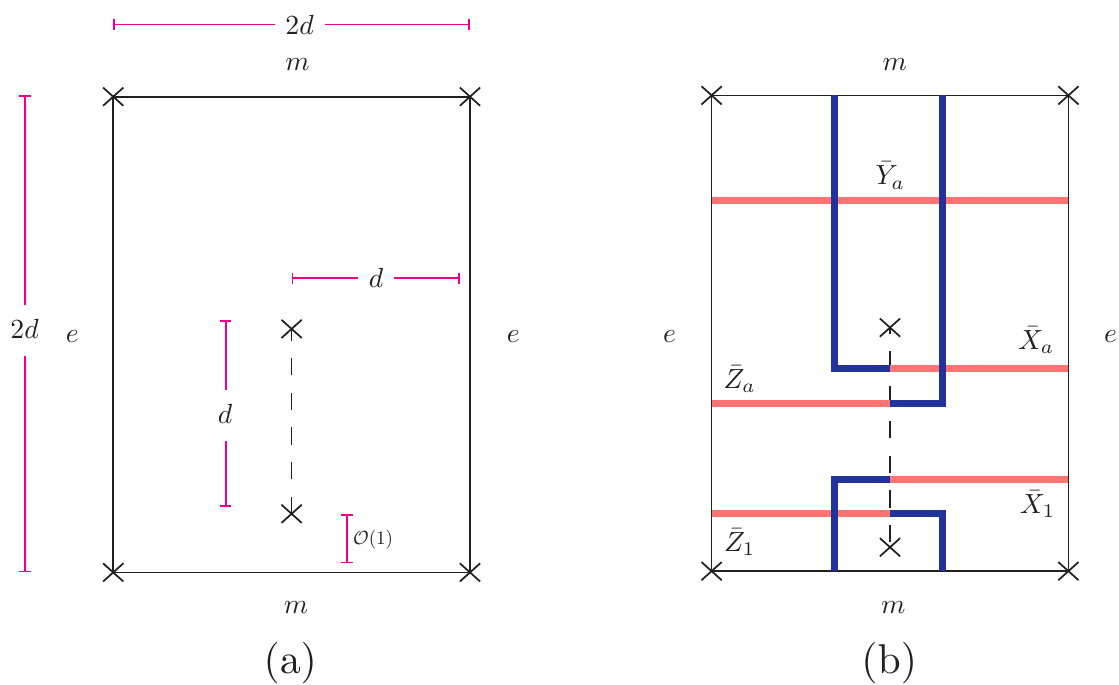}}
\caption{ a) The general layout used to encode a logical data qubit and a logical ancilla qubit with distance $d$ that enables one to do all single qubit Clifford gates explicitly. b) The Pauli strings corresponding to logical operators of the main qubit ($\bar X_1$ and $\bar Z_1$) and the ancilla qubit ($\bar X_a$,$\bar Y_a$ and $\bar Z_a$). }\label{fig:general2}
\end{figure}

\begin{figure}[h]
\centerline{\includegraphics[width=0.3\textwidth]{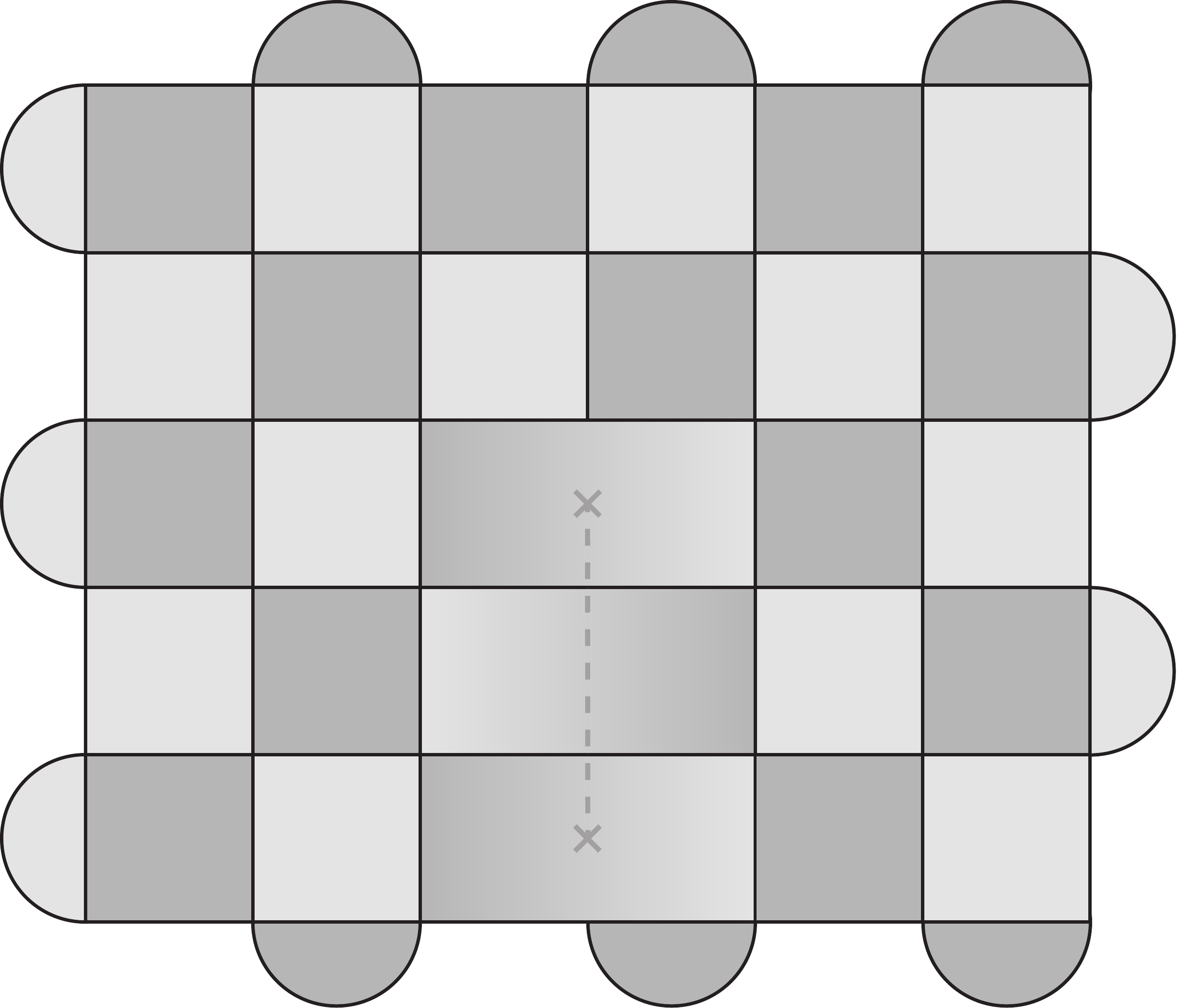}}
\caption{Distance $d=3$ surface code encoding a logical qubit and an ancilla qubit. Logical operators are defined according to Fig. \ref{fig:general2}. }\label{fig:2qubitLattice}
\end{figure}

A minimal setup that allows one to implement all single qubit Clifford gates using the joint measurement
methods of this paper is shown in Fig. \ref{fig:general2}a. In this configuration, the
logical data and ancilla qubits are encoded with both boundary and bulk twist defects. For large $d$, this
configuration requires $\sim 4d^2$ physical qubits. The explicit $d=3$ lattice construction
corresponding to this configuration is shown in Fig. \ref{fig:2qubitLattice}.

\begin{table*}
\centering
\begin{tabular}{l | c c c |c}
\hline
\multicolumn{1}{c |}{$d$}                  & $3$  & $5$  & $7$   & $d$           \\ \hline  \hline

Surface Code (surgery / without distillation)                     & $27$ & $75$ & $147$  & $\sim 3\,d^2$      \\
Surface Code (surgery / one round of distillation)                & $55$ & $138$ & $259$  & $\sim 4.75 \,d^2$      \\
Surface Code (surgery / two rounds of distillation)               & $76$ & $271$ & $343$  & $\sim 6.06\, d^2$      \\
Surface Code (Triangular)                                         & $37$ & $91$ & $169$ & $\sim 3\,d^2$   \\
{\bf Surface Code (Joint Measurement) }                                 & $40$ & $106$ & $204$ & $\sim 4\,d^2$
\end{tabular}
\caption{The number of physical qubits needed in order to implement all single qubit Clifford gates in various schemes. We explain
the calculation of the overhead costs for the state distillation protocols in Appendix B. }\label{tbl:qovh2}
\end{table*}

In Table \ref{tbl:qovh2}, we list the number of physical qubits each design uses to implement all single qubit Clifford gates.
The syndrome qubits which are used for the stabilizer measurements are not included in any of the counting. In case of
the triangular code, three different schemes have been proposed for implementing single qubit Clifford gates\cite{yoder2017};
here we used the code conversion (CC) approach for comparison since in this approach single qubit gates are actually
implemented rather than kept track of classically. Furthermore, a special design of the triangular code for $d=3$ is
proposed in Ref. \onlinecite{yoder2017} which uses only 7 qubits. However, since it cannot be generalized to larger
code distances, it is not included in this comparison.

In the case of the surface code proposal using the planar encoding with lattice surgery, we have included the number of physical qubits
needed for implementing the full Clifford gate set using zero, one, and two rounds of state distillation. Higher numbers of
distillation rounds exponentially decreases the error probability in the purified state.
One round of distillation using the Steane code\cite{steane1996} uses $7$ instances of noisy logical qubits
in the $\ket{Y}$ state to generate a less noisy $\ket{Y}$ state \cite{fowler2012}. If the input state has error probability $p$, the
output state will have $7p^3$ probability of having error. So, performing $k$ rounds of state distillation to reduce
the error probability to $7^{(3^k-1)/2}p^{3^k}$, needs $7^k$ extra logical ancilla qubits (However they need not be prepared with the
full code distance. See Appendix B). Apart from the noisy ancilla qubits initialized in the $\ket{Y}$ state, the distillation process
has a number of other overhead costs as well. These include other ancilla qubits initialized in the standard computational basis
and additional ancilla qubits required in the planar code layout to perform the logical gate operations with lattice surgery \cite{horsman2012}.
The latter for example raises the overhead to $4 \times 7^k$ logical ancilla qubits.
The numbers for one and two rounds of distillation in Table \ref{tbl:qovh2} only include the number
of physical qubits one needs to prepare the initial noisy $\ket{Y}$ states, thus not including these additional resource costs.

The necessary number of distillation rounds depends on the required accuracy of the purified state. If we assume
preparation error and storage error are of the same order, it is reasonable to perform $k\sim \mathcal{O}(\log d)$
rounds of error correction to keep the logical error probability at $\mathcal{O}(p^{d/2})$. This implies that the asymptotic
scaling of the space overhead for protocols that require state distillation is $\mathcal{O}(d^3)$.

\subsection{Two qubit codes}

Now we consider small codes that allow full implementation of the Clifford group on two logical qubits, which thus includes the CNOT gate.
Almost all proposed designs use the circuit shown in Fig. \ref{fig:CNOTgate} for performing CNOT, which requires an additional logical ancilla qubit.
Two important exceptions are the braiding of bulk defects and transversal methods. Braiding methods need a large space to move defects around
and are not suitable for small numbers of qubits. For example the qubit overhead for the double hole implementation scales like $\sim 37/2 d^2$,
which is much worse than other methods\cite{horsman2012} and will not be considered here. The transversal methods, in contrast,
allow implementation of logical CNOT by independent application of CNOT among physical qubits from different code patches. This
uses the minimum number of qubits by eliminating the need for any logical ancilla qubit. However in a two-dimensional single-layer planar geometry, this
method relies on long-range interactions when the system scales up to large numbers of qubits and large code distance $d$. We
restrict our comparison to methods that utilize only local interactions on a planar geometry in this limit, hence omitting transversal methods from the comparison.

\begin{figure}[h]
\centerline{\includegraphics[width=0.4\textwidth]{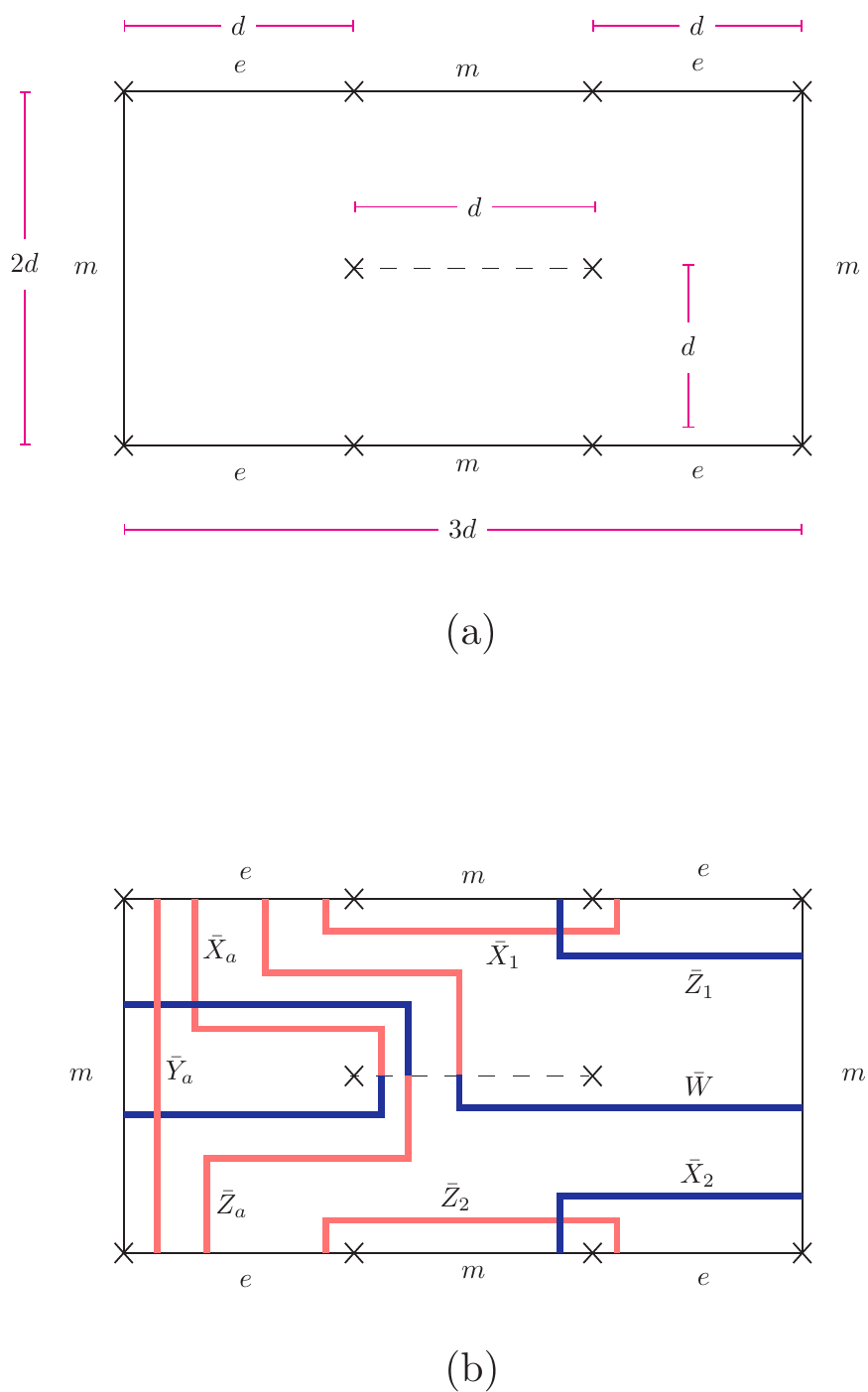}}
\caption{a) The general layout used to encode two logical data qubits and a logical ancilla qubit that enables one to do all Clifford gates explicitly. b) The Pauli strings corresponding to logical operations. The one belonging to the ancilla is specified by $a$ subscript. The $\bar W$ string is an idle string that should be initialized to $+1$ so one can use it for fault tolerant joint measurements.}\label{fig:general3}
\end{figure}

\begin{figure}[h]
\centerline{\includegraphics[width=0.4\textwidth]{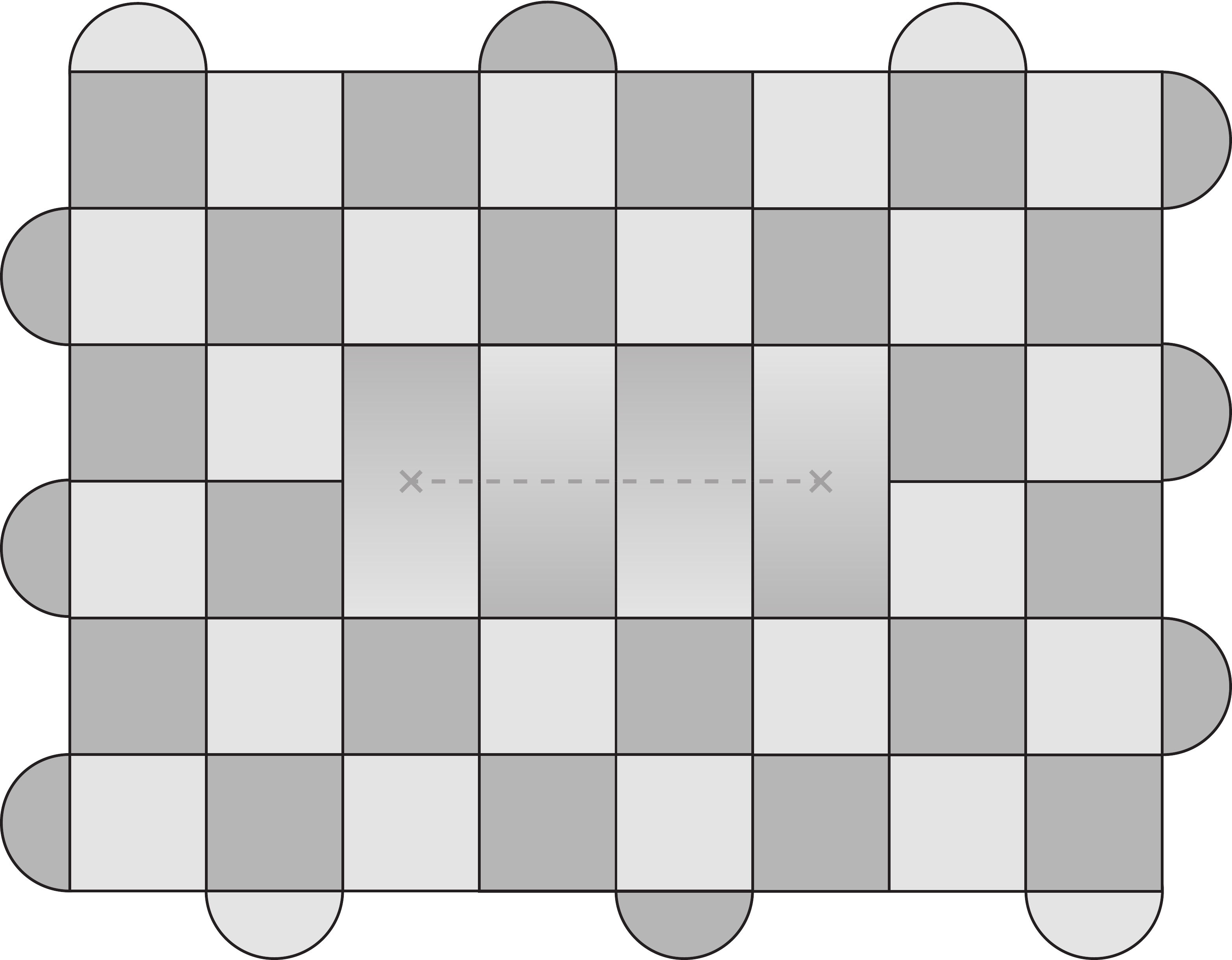}}
\caption{Distance $d=3$ surface code encoding two logical qubit and one ancilla qubit. Logical operators are defined according to Fig. \ref{fig:general3}.}\label{fig:3qubitLattice}
\end{figure}

Fig. \ref{fig:general3}a is a minimal surface code configuration that can be used to encode two logical data qubits and one logical ancilla qubit,
which allows us to implement all Clifford gates using the joint measurement protocols.  The ancilla qubit is encoded with twist defects
and the logical qubits are encoded using boundary defects. For large $d$, this construction uses $\sim 6d^2$ physical qubits.
The explicit $d=3$ lattice construction corresponding to this configuration is also shown in Fig. \ref{fig:3qubitLattice}.



\begin{table*}[t]
\centering
\begin{tabular}{l | c c c c}
\hline
\multicolumn{1}{c |}{$d$}                  & $3$  & $5$  & $7$   & $d$           \\ \hline  \hline
Surface Code (surgery / without distillation)                     & $36$ & $100$ & $196$  & $\sim 4\,d^2$      \\
Surface Code (surgery / one round of distillation)                & $64$ & $163$ & $308$  & $\sim 5.75\, d^2$      \\
Surface Code (surgery / two rounds of distillation)               & $85$ & $296$ & $392$  & $\sim 7.06\, d^2$      \\
Surface Code (Triangular)        & $74$ & $182$ & $338$ & $\sim 6\,d^2$   \\
{\bf Surface Code (Joint Measurement)} & $60$ & $160$ & $308$ & $\sim 6\,d^2$   \\

\end{tabular}
\caption{The number of physical qubits to implement all Clifford group gates including CNOT. We use the results of Appendix B for the state distillation overhead costs.}\label{tbl:qovh3}
\end{table*}

Table \ref{tbl:qovh3} lists the number of physical qubits that various methods use to implement the full Clifford group. Again, the numbers for surface code with distillation reflect only the number of qubits used in encoding noisy $\ket{Y}$ states in the lowest layer of distillation. As one can see, the joint measurement method of this paper uses the minimum number of physical qubits among
other surface code variants, but scales the same as the triangular code for large $d$.

\subsection{Many qubit codes}\label{sec:manyqubits}

The two previous sections discussed encodings for one or two logical qubits (and a logical ancilla). If we consider large
scale quantum computing with more than just three total qubits, there are more clear advantages for using the proposed joint measurement
design for gates.

First, the joint measurement protocol naturally allows CNOT gates between two logical qubits that are arbitrarily far apart from each other.
By encoding all logical qubits within the same patch of surface code (e.g. using bulk twist defects or holes), we can measure arbitrarily long
strings with no additional space-time overhead. This means that even the logical ancilla qubit in the CNOT circuit does not need to be physically
near the control and target logical qubits. On the other hand, the lattice surgery based methods (which is used by both planar and triangular
surface codes) only allow operations between adjacent patches of the code. Thus, a CNOT gate between far-separated qubits requires bringing
the two patches near each other first. Apart from the time overhead this operation imposes, it necessitates sufficient blank patches
between data qubits, which results in increased space overhead; One can show that for a typical arrangement of square patches of
planar codes, only a quarter of patches can be used for storing data and other patches should be reserved as blank spaces to be used
for moving data patches around\cite{horsman2012}.

A natural question now is which particular type of encoding is most efficient, in the limit of large numbers of logical qubits, for minimizing
the number of physical qubits for a given code distance. We find that the optimal encoding is with a lattice of bulk twist defects (dislocations).
Such a dislocation code was discussed in Ref. \onlinecite{hastings2014}; however our proposal and results for resource estimates differ somewhat
from those reported previously. Specifically, we consider an arrangement of twist defects as shown in Fig. \ref{fig:dislocationarrangement}.
The twist defects are placed on a rotated square lattice with lattice constant $d/\sqrt{2}$, in contrast to the square lattice with lattice constant $d/2$ that
was proposed in Ref. \onlinecite{hastings2014}. We require this modification to protect the code from Pauli-$Y$ error strings that can start and end
on the twist defects.

Furthermore, since long-range CNOT gates require measuring long-range string operators, it is important to keep the code distance $d$ throughout the
measurement process. To ensure this, we initialize the idle strings (like $\bar W$ in Fig.~\ref{fig:dislocationarrangement}) to $+1$ and by utilizing them,
we thread the strings through the available space between the defects in such a way that every twist defect is at least distance $d/4$ apart from the
measured string, as is shown in Fig. \ref{fig:dislocationarrangement}. This ensures that no short error string could happen while measuring the strings.
More details can be found in the caption. By adding one ancilla qubit to the patch and using the joint measurement method, one can apply all Clifford
gates fault tolerantly.

In Table \ref{tbl:qovhN}, we have listed encoding rates for various schemes for comparison. For the lattice surgery method on planar codes, the $(7/4)^k~d^2$ term
arises from the additional logical ancilla qubits needed for state distillation(See Appendix \ref{distapx}). As was explained in Sec. \ref{sec:singlequbit}, one needs
$\mathcal{O}(\log d)$ rounds of state distillation to obtain $\sim p^{d/2}$ logical error probability. This in turn means the additional cost due to state
distillation grows like $d^3$ and dominates the resource usage. In the case of the triangular code, Ref. \onlinecite{yoder2017} proposed several protocols;
in Table \ref{tbl:qovhN}, we quoted the encoding rate when one keeps track of single qubit gates at the classical level instead of applying them
directly on the quantum code (referred to as the basis-state conversion (BC) scheme). We see that the joint measurement method
with a lattice of bulk twist defects performs better than both of these.

We note that the direct comparison of different encoding schemes summarized in Table \ref{tbl:qovhN} involves some subtleties as the different schemes also
have various relative advantages. For example, the joint measurement estimate assumes a single logical ancilla for an arbitrary number of logical gates,
while the estimates for the other schemes assume a number of logical ancillas that grows with the number of logical qubits. On the other hand,
the joint measurement scheme allows arbitrarily long-range CNOT gates, while the other two schemes do not. Furthermore, we note that Ref. \onlinecite{litinski2017}
has proposed an alternative patch-based scheme that allows classical tracking of the single-qubit Clifford gates; however a direct comparison with
that proposal is more complicated, as it allows for many possible distinct designs.

\begin{table}
\vspace{5pt}
\centering
\begin{tabular}{l | c }
\hline
Scheme                           & $N_{phys}/N_{L}$ \\ \hline  \hline
Surface Code (Surgery / $k$ rounds of distillation)           & $4(1+\qty(\frac{7}{4})^k)\,d^2 \sim d^3 $  \\
Surface Code (Triangular)        & $9/4\,d^2$  \\
{\bf Surface Code (Joint Measurement)} & $3/2\,d^2$  \\
\end{tabular}
\caption{Asymptotic encoding ratio for large number of logical qubits, while being able to implement all Clifford gates. }\label{tbl:qovhN}
\end{table}

One can improve these scaling relations for the encoding rate by using non-local interactions and thus implementing hyperbolic codes. As discussed in Sec. \ref{sec:hyper},
our joint measurement proposal is so far the only proposed method to implement the full Clifford group in hyperbolic codes.

\section{Conclusion}
\label{sec:conclusion}

We have shown that every encoding scheme in the surface code admits a fault-tolerant implementation of the full Clifford gate set, without the need for
state distillation or conversion between different types of encodings as required in a number of previous proposals.
If the logical ancilla is encoded using boundary defects, holes, or through non-trivial genus, then a CAT state can be used to implement
the required topological charge measurements. On the other hand, if the logical ancilla is encoded with bulk twist defects, then the CAT state is not necessary,
and the full Clifford gate set can be directly implemented. We have further shown how these methods allow implementations of the Clifford group in the
3D surface code with a sphere encoding. We also used these methods to provide the first proposals for implementing the full Clifford group in hyperbolic codes,
and a new scheme for implementing them in the color code, which allows arbitrarily long-range CNOT gates.

Our joint measurement proposal also informs efficient surface code designs. In particular, we have proposed designs for one and two logical qubits
and one logical ancilla qubit, which admit fault-tolerant implementation of the full Clifford group. In the case of two logical qubits and one ancilla,
our proposal is an improvement over all previous proposals in terms of number of physical qubits required for a given code distance $d$. As such,
they may be of use for near-term experiments to demonstrate fault-tolerant implementation of logical gates.

For large numbers of logical qubits, we have found that a lattice of twist defects (the dislocation code) is optimal in terms of spatial overhead.
While such a code was studied in Ref. \onlinecite{hastings2014}, our analysis of the resource overhead and scheme for performing long-range
CNOT gates is distinct.

Ultimately, for quantum computing applications, an added advantage of the dislocation code, as pointed out in Ref. \onlinecite{hastings2014}, is
that the single-qubit Clifford gates add nothing to the time overhead, as they can effectively be absorbed into the CNOT gates by picking different
strings to measure along. Similar advantages exist in the triangular code\cite{yoder2017} and certain patch-based proposals, as
discussed in Ref. \onlinecite{litinski2017}. We have shown that similar advantages exist for our protocols as well and thus exist for
any encoding scheme, as long as logical ancilla qubits are implemented in the surface code with twist defects. In contrast, the single qubit
Clifford gates will need to be performed for the proposals of Ref. \onlinecite{fowler2012,horsman2012}. Nevertheless, it presumably
would be useful to actually implement single-qubit Clifford gates in small near-term quantum computers in order to demonstrate
experimentally the possibility of implementing fault-tolerant logical gates.

\section{Acknowledgment}

We thank Isaac Kim for helpful discussions. This work is supported by
NSF CAREER (DMR-1753240) and JQI-PFC-UMD.

\appendix
\section{Time overhead estimate for CAT state measurements}\label{catapx}

Here we discuss the time overhead for utilizing CAT states for performing logical measurements.
In order to measure a string operator $S=\Pi_{i=1}^d\sigma_i$ using cat states, first we prepare the
measurement qubits in the $\ket{+}^d$ state and then measure $Z_i Z_{i+1}$ stabilizers for $d$ rounds.
Next we perform controlled $\sigma_i$ gates between the measurement qubits and corresponding data qubits
and at last we measure measurement qubits in $X_i$ basis and infer the measurement results form the overall parity of outcome.

Bit flip errors are generally taken care of by the syndrome measurements so here, we concentrate on the phase flip errors.
Since phase flip errors cannot be detected by stabilizer measurements, we need to repeat the measurement many times
and decide on the value of measurement by taking a majority vote. Here we provide an estimate on the number of measurements $r$
one needs to achieve fault tolerance.

We only consider phase flip errors that occur during storage intervals. It is reasonable to expect that including other
types of errors will not change the answer significantly.

We have $d$ measurement qubits. During preparation of the CAT state, we perform $d$ rounds of stabilizer measurements.
So during preparation, we consider $d^2$ space-time points where phase flip errors can happen. Applying controlled
$\sigma_i$ gates can be done simultaneously in one step, therefore it adds $d$ places to the potential locations for phase
flip errors. Thus there are $\mathcal{O}(d^2)$ spots where phase flip errors can happen during a single round of CAT state measurement.
For simplicity we take this number to be $d^2$.

Let's say the probability of a phase flip error affecting a single qubit during a single time step is given by $p$. Note that
if an \textit{even} number of such errors occur during a round of measurement, we still get the right answer. To find the
probability of this happening, we use the following recursive expression for having even number of errors in the first $n$ locations,
\begin{equation}
  P_{n+1}=P_n(1-p)+(1-P_n)p.
\end{equation}
This difference equation can be solved easily to get,
\begin{equation}
  q\equiv P_{d^2}=\frac{1}{2}+\frac{1}{2}(1-2p)^{d^2}.
\end{equation}
Although this result is exact, we are only interested in the $p\ll 1$ limit:
\begin{equation}
  q=\frac{1}{2}+\frac{1}{2}e^{-2p\,d^2}
\end{equation}

$q$ gives the probability of getting the right answer after a single measurement. If $pd^2\ll 1$, then we have,
\begin{equation}
  q=1-pd^2,
\end{equation}
and therefore to reduce the probability of failure to $p^{d/2}$, it suffices to repeat the measurement $\mathcal{O}(d)$ times,
\begin{equation}
  r\sim \mathcal{O}(d), \qquad pd^2\ll 1.
\end{equation}
However, if $pd^2\gg 1$, $q$ will be exponentially close to $1/2$ and we need a huge number of measurements to make sure the majority vote gives the right answer.

To find out how many times we need to repeat the measurement, we can use the central limit theorem. First we define the following statistical variable:
\begin{equation}
  X=
  \begin{cases}
    1&\text{with probability } q\\
    0&\text{with probability } 1-q
  \end{cases}
\end{equation}
basically $X$ would be $1$ when we get the right answer and $0$ otherwise. Note that the mean and variance of $X$ is given by $\expval{X}=q$ and $\sigma^2_X=q(1-q)$ respectively. If we repeat the measurements $r$ times, the quantity which we are interested in is given by,
\begin{equation}
  S = \frac{\sum_{i=1}^r X_i}{r}
\end{equation}
If $S>1/2$ the majority vote gives the right answer and if $S<1/2$ we get the wrong answer. It is well known that for $r\gg1$, the distribution function of $S$ can be well approximated by a normal distribution around $\mu_S=\expval{X_i}$ with standard deviation $\sigma_S=\frac{\sigma_X}{\sqrt{r}}$.

Using this distribution, it is easy to see that the probability of getting $S>1/2$ is given by:
\begin{equation}\label{pf}
  P_{\text{failure}}=\frac{1}{2}\,\erfc(e^{-2p d^2}\sqrt{\frac{r}{2}})
\end{equation}
where $\erfc(x)$ is the complementary error function given by,
\begin{equation}
  \erfc(x)=\frac{2}{\sqrt{\pi}}\int_x^{\infty}e^{-t^2}\dd t.
\end{equation}
For large $x$s, $\erfc(x)$ can be approximated by,
\begin{equation}
  \erfc(x)=\frac{e^{-x^2}}{\sqrt{\pi}x}(1+\mathcal{O}(x^{-1}))
\end{equation}
using this asymptotic form alongside Eq.\eqref{pf}, one can see that for $pd^2\ll 1$, to get $P_{\text{failure}}\sim p^d$, we need to repeat the measurement exponentially many times,
\begin{equation}
  r\sim \mathcal{O}(e^{\alpha d^2}),\qquad pd^2\gg1.
\end{equation}
We remark that although this will make the method fault tolerant, it renders this method quite unpractical in this regime.

\section{Space overhead for state distillation}\label{distapx}
\begin{figure}
  \centering\includegraphics[width=0.9\columnwidth]{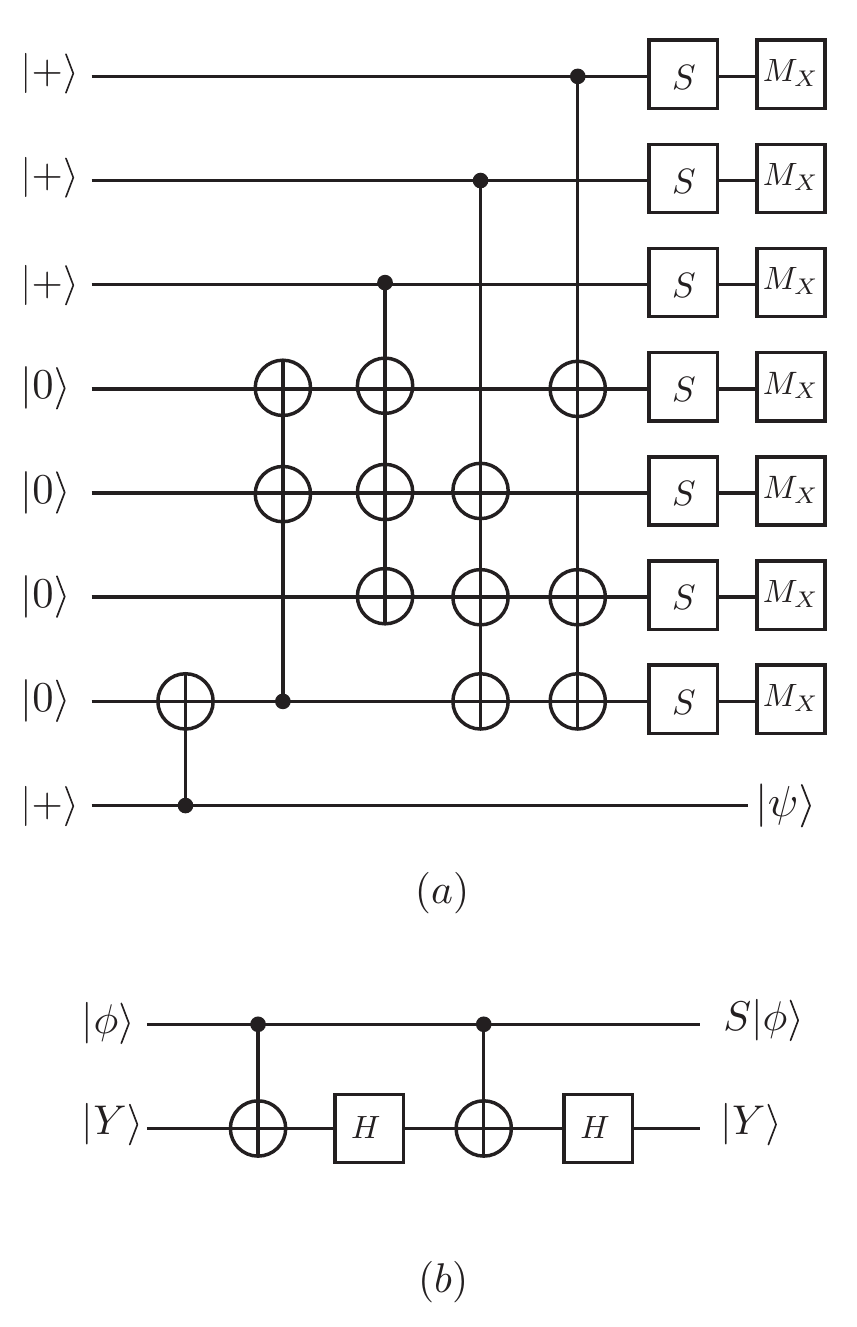}
  \caption{a)The quantum circuit used to purify $\ket{Y}$ states. b) The quantum circuit used to implement $S$ gates in the distillation circuit which uses an ancilla prepared in $\ket{Y}$ state. The ancilla is either initialized by state injection technique(in the lowest layer of distillation) or is the purified output of previous rounds of distillation.}
  \label{fig:dist}
\end{figure}

In this section we briefly estimate the space overhead incurred by state distillation and thus explain how the numbers in Tables \ref{tbl:qovh2} and \ref{tbl:qovh3} were calculated.

The distillation circuit for logical $\ket{Y}$ state based on the Steane code\cite{steane1996} is illustrated in Fig.\ref{fig:dist}a\cite{fowler2012}.
We will not cover the details of this circuit here; the interested reader can consult Refs.\onlinecite{fowler2012,steane1996} for a through
explanation. Basically, in each round of distillation seven noisy $\ket{Y}$ states are input into the circuit through the $S$ gates
(as shown in Fig.\ref{fig:dist}b\cite{fowler2012}) and after the final measurements, with some finite fixed probability,
$\ket{\psi_L}$ will collapse to a purified version of $\ket{Y}$ (or $\bar Z \ket{Y}$ which can be corrected easily).

For $k$ rounds of distillation, one starts with $7^k$ noisy logical qubits prepared in the $\ket{Y}$ state using state injection. After the first round
of distillation, $7^{k-1}$ purified states would be produced, which then will be used for the next round of distillation. Applying distillation $k-1$
more times by using the output of previous rounds as the input to the next round will result in a single distilled $\ket{Y}$ state with error
probability $\sim p^{3^k}$. A lower bound for the space overhead of distillation can be obtained by only counting the number
of qubits used in preparing those $7^k$ initial $\ket{Y}$ states.

When performing many layers of distillation, there is no point in using logical qubits with the full distance $d$ in every layer, as required in the final
layer of distillation, since the input states are noisy anyways. Instead one can, for example, halve the code distance at each lower level to reduce the
space overhead. Considering this, initial $\ket{Y}$ states can be prepared using surface codes of distance $\frac{d}{2^k}$. If $\frac{d}{2^k}<1$ then there is no need to encode the initial states into the surface code by state injection; one can just start with $7^k$ physical qubits initialized in $\ket{Y}$ state.

Therefore in this scheme we need a total of
\begin{equation}
  N=7^k\times \Big\lceil \frac{d}{2^k} \Big\rceil^2
\end{equation}
physical qubits just to encode the noisy $\ket{Y}$ states in the lowest level.  Note that this is a lower bound on the overhead incurred by state distillation,
since we have not included other ancilla qubits which are used in distillation circuit. The numbers in Tables \ref{tbl:qovh2} and \ref{tbl:qovh3}
for surface code with distillation are calculated by adding this overhead to basic resource costs. We also note that it may be possible to further optimize the state distillation
protocol and thus further reduce the constant factors involved in the space overhead; a complete analysis of such an optimization is beyond the scope of this paper. 
\bibliography{TI}

\end{document}